\documentclass[12pt]{article}

\usepackage{indentfirst}
\usepackage{titlesec}
\titleformat*{\section}{\large\bfseries}
\titleformat*{\subsection}{\normalsize\bfseries}

\usepackage[symbol]{footmisc} 
\renewcommand*{\thefootnote}{\fnsymbol{footnote}} 

\usepackage[colorlinks]{hyperref}
\hypersetup{linkcolor=blue,citecolor=blue,urlcolor=blue}

\usepackage{amsmath}
\usepackage{amssymb}

\usepackage{physics}

\usepackage{graphicx,color}
\usepackage{xcolor}

\usepackage{float} 

\usepackage{axodraw2} 
\usepackage[compat=1.1.0]{tikz-feynman}

\usepackage{comment} 
\usepackage[normalem]{ulem} 

\def\nn{\nonumber}
\def\beq{\begin{eqnarray}}
\def\eeq{\end{eqnarray}}

\def\ln{\,\mbox{ln}\,}

\def\Tr{\,\mbox{Tr}\,}

\def\al{\alpha}
\def\be{\beta}

\def\ga{\gamma}

\def\ep{\epsilon}

\def\la{\lambda}
\def\na{\nabla}
\def\pa{\partial}

\def\si{\sigma}
\def\om{\omega}
\def\ph{\varphi}

\def\Ga{\Gamma}

\def\cl{\mathcal}

\def\tx{\text}

\usepackage[backend=biber, sorting=none, url=false, doi=false, isbn=false,
giveninits=true, uniquename=init]{biblatex}
\begin{document}

\begin{center}
	{\Large \bf
    Covariant One-Loop Corrections from a Non-Covariant Expansion:
		\vskip 2mm
		The Scalar Field Case}
\end{center}

	\vskip 6mm

\begin{center}	
	{\bf Lucas Ducastelo de C. C. Lima,}$^{a}$
	\hspace{-1mm}\footnote{E-mail address: \ ldcc.lima@unesp.br}
	\hspace{0.3cm}
	{\bf Riccardo Fecchio}$^{b,c}$
	\hspace{-1mm}\footnote{E-mail address: \ riccardo.fecchio@edu.ufes.br}
	
	\vskip 6mm
	$^{a}$ Programa de P\'os-Gradua\c{c}\~ao em F\'{\i}sica e Astronomia,
	\\
	Universidade Estadual Paulista (Unesp), Guaratinguet\'a, 12516-410,  S\~ao Paulo, Brazil.
	\vskip 2mm
	
	$^{b}$ 	PPGCosmo, Universidade Federal do Esp\'{\i}rito Santo
	\\
	Vit\'oria, 29075-910, ES, Brazil
	\vskip 2mm
	
	$^{c}$ Departamento de F\'{\i}sica, ICE, Universidade Federal de Juiz de Fora,
	\\
	Juiz de Fora, 36036-900, Minas Gerais, Brazil
	\vskip 2mm
	
\end{center}

\vskip 2cm

\begin{abstract}
	We compute the one-loop correction to the non-minimal coupling of a scalar field theory to gravity, induced by a quartic self-interaction. We adopt the non-covariant framework of linearized gravity and explicitly compute the corresponding Feynman diagram, including both the divergent and the finite parts.
	The corresponding contribution to the one-loop effective action is obtained by the requirement of covariance and compared with the known results obtained by the heat kernel technique. 
	We study the running of the non-minimal coupling, its UV and IR limit and verify the validity of the decoupling theorem in the momentum subtraction scheme of renormalization.
	The linearized gravity approach is also used for deriving the effective potential for the scalar field and its covariant version is stated.
	The agreement with results derived using strictly covariant methods is verified and highlights the versatility of the linearized gravity approach. 
\end{abstract}

\setcounter{footnote}{0} 
\renewcommand*{\thefootnote}{\arabic{footnote}} 
\newpage

\section{Introduction}
\label{sec1}

The Effective Action is an important objects for understanding quantum effects in field theory. 
It may be obtained in the framework of the background field method, where the aim is to compute the dynamics of a background field taking into account quantum fluctuations.
The dynamics of the background field is encoded into the effective action that, due to quantum corrections, usually shows non-local form factors.
An important result from this approach is the Weinberg-Coleman potential, i.e.\ the effective potential for a scalar field, which presents the logarithmic correction \cite{Coleman:1973jx}. 

An expected property of the quantum corrections is a smooth low-energy decoupling for massive degrees of freedom. This property was initially stated in the Appelquist-Carazzone theorem \cite{Appelquist:1974tg}, first formulated in QED.
The form factors in the effective action are crucial for understanding the physical, scale-dependent transitions of the theory from the ultraviolet (UV) to the infrared (IR) regimes and verify the decoupling.

	In curved spacetime, the calculation of the effective action is more involved and usually is performed by using covariant techniques.
	Nowadays, the heat-kernel method offers an elegant and manifestly covariant framework to derive the effective action and it is widely used for studying one-loop physics in both semiclassical and quantum gravity \cite{Barvinsky:1985an, Avramidi:1986mj, DeWitt:2003pm}.

	The resulting effective actions present non-local form factors \cite{BirrellDavies, Buchbinder:1985ew, BuchbinderShapiro:2021} and the decoupling property is respected in curved spacetime too \cite{Gorbar:2003I, Gorbar:2003II, Borges:2025gfg}.
The gravitational form factors and the mechanism of matter field decoupling in four-dimensional gravity were further investigated in \cite{Franchino-Vinas:2018gzr, Mottola:2026vacuum} (see \cite{Ribeiro:2018pyo} for the two-dimensional case).

Although the heat-kernel method is computationally efficient, an independent diagrammatic calculation in linearized gravity offers an essential check of its results. A check of this kind was carried out in \cite{Gorbar:2003I}, where the heat-kernel result for the vacuum effective action induced by a massive scalar field was reproduced diagrammatically.
In the present work, we use linearized gravity to derive the one-loop effective action of a scalar field, thereby providing an independent check of the heat-kernel result \cite{deBerredo-Peixoto:2003pkj}.

The linearized gravity approach is based on expanding the metric into a flat background plus a \emph{classical} fluctuation $h_{\mu\nu}$ and computing the quantum corrections in powers of $h_{\mu\nu}$.
Because of the metric expansion, covariant symmetry is explicitly broken and intermediate results are non-covariant. However, once all the orders in the metric fluctuations are summed up, the covariant symmetry is restored and covariant results are expected. 
Because of the non-linearity of gravity, there are infinitely many orders in $h_{\mu\nu}$. This could seem a limitation to the applicability of the linearized gravity approach but it is not, since one can eventually extrapolate the final, covariant results from the first few orders in $h_{\mu\nu}$. 
Moreover, the expansion around a flat spacetime allows to elaborate expressions in momentum space. 

In this work we perform the explicit calculation of the one-loop correction to the non-minimal coupling $\xi$ for a scalar field in curved spacetime with quartic self-interaction, i.e.\ we consider the theory (in Euclidean signature)
\begin{align} \label{eq:action-Euclidean}
\mathcal{S}\,=\, \int \dd[4]{x}\sqrt{g}\left\{\frac{1}{2}\, g^{\mu\nu} \na_\mu \ph \na_\nu \ph + \frac{1}{2}m^2\ph^2 +\frac{1}{2}\xi R\ph^2 + \frac{\lambda }{4!}\ph^4 \right\} \,.
\end{align}
We derive both the one-loop effective action for the $\xi$-sector and its beta function. Furthermore, we compute the effective potential and discuss the limits in which the effective action reduces to the effective potential.

We extrapolate the covariant correction to the effective action using the linearized gravity approach at first order and check the consistency of our result with the one derived by heat-kernel \cite{deBerredo-Peixoto:2003pkj}.
The effective potential is computed up to second order in linearized gravity, and we comment on why this specific order is necessary to extract the correct covariant result.
At each step, a diagrammatic interpretation of formulas is given, based on Feynman diagrams. 
Throughout the paper, divergences arising from quantum loops are regularized by dimensional regularization and renormalization is discussed. 

The agreement between the linearized gravity approach and covariant approaches such as the heat-kernel method and
the local momentum representation based on normal coordinates (although the latter captures only local corrections \cite{BuchbinderShapiro:2021}) shows the versatility of the method, which in some situations could provide an easier way of extracting quantum corrections in curved spacetime.
Perhaps more importantly, this method can be extended to higher-loop orders and to fields other than the scalar, at the price of more involved calculations. By contrast, standard heat-kernel techniques are primarily tailored to one-loop calculations, whereas diagrammatic methods provide a more
direct framework for extensions beyond one loop.

The article is organized as follows. 
In Section~\ref{sec:2} we give an overview of one-loop quantum corrections to a scalar field in linearized gravity, where any order in the gravitational field is treated as an external tail in the corresponding Feynman diagrams. 
In Section~\ref{sec:3} we focus on the ``tadpole'' diagram\footnote{This diagram is a tadpole only in flat spacetime. Once gravity is turned on, external momentum enters the loop, as detailed later in the text.} producing the running of the non-minimal parameter $\xi$ and discuss the divergent and finite parts of the diagram. Also, we comment on the covariantization of the diagrammatic result as well as its UV and IR limits.
We conclude the section discussing the beta function for $\xi$.
To the best of our knowledge, the tadpole diagram we compute in Section~\ref{sec:3} has never been explicitly evaluated before.
In Section~\ref{sec:4}, we construct a covariant effective potential starting from the framework of linearized gravity. Finally, in Section~\ref{sec:5}, we draw our conclusions.

\section{Overview of one-loop corrections for a scalar field}
\label{sec:2}
In curved spacetime, the most general covariant action for a real scalar field, with coupling constants of non-negative mass dimension, is given by eq.\ \eqref{eq:action-Euclidean}. There, $\xi R \ph^2$ is the non-minimal term, having no analogue in flat spacetime. 
Theory \eqref{eq:action-Euclidean} is perturbatively renormalizable. Quantum effects were studied using the heat-kernel technique in \cite{deBerredo-Peixoto:2003pkj}.
Equivalently, the diagrammatic approach can be used. The approach is based on considering the classical metric $g_{\mu\nu}$ as an external field and building Feynman diagrams in a flat spacetime by perturbing the action for $g_{\mu\nu}=\delta_{\mu\nu}+h_{\mu\nu}$, where $\delta_{\mu\nu}$ is the flat (Euclidean) background and $h_{\mu\nu}$ is an external, perturbation field. The advantage of the diagrammatic approach is to have diagrams in a flat background that can be solved in momentum space as usual in QFT. On the other hand, the price paid is the breaking of covariance, due to the metric expansion. A covariant result is obtained summing up all the diagrams, to all orders in $h_{\mu\nu}$, of which there are infinitely many.
However, from power counting and dimensionality, one grasps the intuition of the covariant result already from a few diagrams at the first orders in $h_{\mu\nu}$.

To illustrate the approach we consider the theory \eqref{eq:action-Euclidean} and compare our results with those of \cite{deBerredo-Peixoto:2003pkj}, obatined by the heat-kernel method. 
Explicit computations will be shown in the next sections. For the rest of the present section we discuss renormalizability of the theory in linearized gravity. 

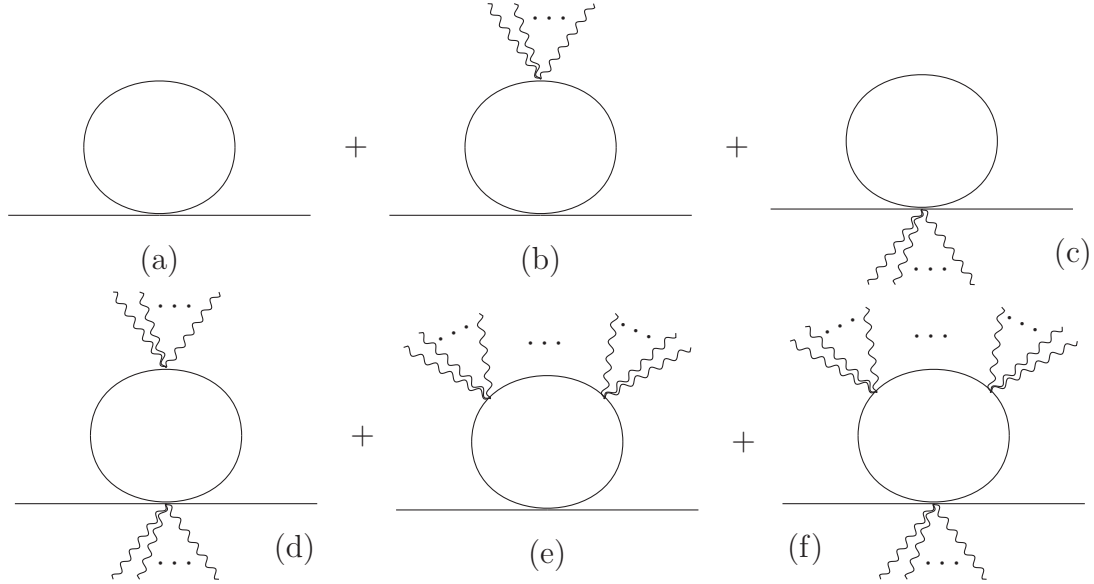
\begin{figure}
	\centering
	\begin{tikzpicture}
		\begin{feynman}
			\vertex (v1) at (0,-0.9);
			\vertex (v2) at (-1,0);
			\vertex (v3) at (1,0);
			\vertex (a1) at (-2,-0.9);
			\vertex (a2) at (2,-0.9);
			\node (name) at (0,-1.5) {(a)};
			\node (plus) at (2.6,0) {+};
			
			\diagram* {
				(a1) -- [plain] (v1),
				(v1) -- [plain] (a2),
				(v2) -- [half left](v3),
				(v2) -- [half right] (v3),
			};
		\end{feynman}
	\end{tikzpicture}
	\begin{tikzpicture}
		\begin{feynman}
			\vertex (v1) at (0,-0.9);
			\vertex (v2) at (-1,0);
			\vertex (v3) at (1,0);
			\vertex (v4) at (0,0.9);
			\vertex (a1) at (-2,-0.9);
			\vertex (a2) at (2,-0.9);
			\vertex (a3) at (-0.7,1.9);
			\vertex (a4) at (0.7,1.9);
			\vertex (a5) at (-0.35,1.9);
			\node (name) at (0,-1.5) {(b)};
			\node (plus) at (2.6,0) {+};
			
			\diagram* {
				(a1) -- [plain] (v1),
				(v1) -- [plain] (a2),
				(v2) -- [half left](v3),
				(v2) -- [half right] (v3),
				(v4) -- [boson] (a3),
				(v4) -- [boson] (a4),
				(v4) -- [boson] (a5),
			};
			\node at (0.15,1.7) {$\cdots$};
		\end{feynman}
	\end{tikzpicture}
	\begin{tikzpicture}
		\begin{feynman}
			\vertex (v1) at (0,-0.9);
			\vertex (v2) at (-1,0);
			\vertex (v3) at (1,0);
			\vertex (a1) at (-2,-0.9);
			\vertex (a2) at (2,-0.9);
			\vertex (a3) at (-0.7,-1.9);
			\vertex (a4) at (0.7,-1.9);
			\vertex (a5) at (-0.35,-1.9);
			\node (name) at (2,-1.5) {(c)};
			
			\diagram* {
				(a1) -- [plain] (v1),
				(v1) -- [plain] (a2),
				(v2) -- [half left](v3),
				(v2) -- [half right] (v3),
				(v1) -- [boson] (a3),
				(v1) -- [boson] (a4),
				(v1) -- [boson] (a5),
			};
			\node at (0.15,-1.7) {$\cdots$};
		\end{feynman}
	\end{tikzpicture}
	\begin{tikzpicture}
		\begin{feynman}
			\vertex (v1) at (0,-0.9);
			\vertex (v2) at (-1,0);
			\vertex (v3) at (1,0);
			\vertex (v4) at (0,0.9);
			\vertex (a1) at (-2,-0.9);
			\vertex (a2) at (2,-0.9);
			\vertex (a3) at (-0.7,1.9);
			\vertex (a4) at (0.7,1.9);
			\vertex (a5) at (-0.35,1.9);
			\vertex (a6) at (-0.7,-1.9);
			\vertex (a7) at (0.7,-1.9);
			\vertex (a8) at (-0.35,-1.9);
			\node (name) at (1.7,-1.5) {(d)};
			\node (plus) at (2.6,0) {+};
			
			\diagram* {
				(a1) -- [plain] (v1),
				(v1) -- [plain] (a2),
				(v2) -- [half left](v3),
				(v2) -- [half right] (v3),
				(v4) -- [boson] (a3),
				(v4) -- [boson] (a4),
				(v4) -- [boson] (a5),
				(v1) -- [boson] (a6),
				(v1) -- [boson] (a7),
				(v1) -- [boson] (a8),
			};
			\node at (0.15,1.7) {$\cdots$};
			\node at (0.15,-1.7) {$\cdots$};
		\end{feynman}
	\end{tikzpicture}
	\begin{tikzpicture}
		\begin{feynman}
			\vertex (v1) at (0,-0.9);
			\vertex (v2) at (-1,0);
			\vertex (v3) at (1,0);
			\vertex (v4) at (0,0.9);
			\vertex (v5) at (-0.75,0.57);
			\vertex (v6) at (0.75,0.57);
			\vertex (a1) at (-2,-0.9);
			\vertex (a2) at (2,-0.9);
			\vertex (a3) at (-1.9,1.25);
			\vertex (a4) at (-1.7,1.44);
			\vertex (a5) at (-0.9,1.7);
			\vertex (a6) at (1.9,1.25);
			\vertex (a7) at (1.7,1.44);
			\vertex (a8) at (0.9,1.7);
			\node (name) at (0,-1.5) {(e)};
			\node (plus) at (2.6,0) {+};
			
			\diagram* {
				(a1) -- [plain] (v1),
				(v1) -- [plain] (a2),
				(v2) -- [half left](v3),
				(v2) -- [half right] (v3),
				(v5) -- [boson] (a3),
				(v5) -- [boson] (a4),
				(v5) -- [boson] (a5),
				(v6) -- [boson] (a6),
				(v6) -- [boson] (a7),
				(v6) -- [boson] (a8),
			};
			\node[rotate=25] at (-1.2,1.45) {$\cdots$};
			\node[rotate=-25] at (1.2,1.45) {$\cdots$};
			\node at (0,1.3) {$\cdots$};
		\end{feynman}
	\end{tikzpicture}
	\begin{tikzpicture}
		\begin{feynman}
			\vertex (v1) at (0,-0.9);
			\vertex (v2) at (-1,0);
			\vertex (v3) at (1,0);
			\vertex (v4) at (0,0.9);
			\vertex (v5) at (-0.75,0.57);
			\vertex (v7) at (0,-0.9);
			\vertex (v6) at (0.75,0.57);
			\vertex (a1) at (-2,-0.9);
			\vertex (a2) at (2,-0.9);
			\vertex (a3) at (-1.9,1.25);
			\vertex (a4) at (-1.7,1.44);
			\vertex (a5) at (-0.9,1.7);
			\vertex (a6) at (1.9,1.25);
			\vertex (a7) at (1.7,1.44);
			\vertex (a8) at (0.9,1.7);
			\vertex (a9) at (-0.7,-1.9);
			\vertex (a10) at (0.7,-1.9);
			\vertex (a11) at (-0.35,-1.9);
			\node (name) at (-1.7,-1.5) {(f)};
			
			\diagram* {
				(a1) -- [plain] (v1),
				(v1) -- [plain] (a2),
				(v2) -- [half left](v3),
				(v2) -- [half right] (v3),
				(v5) -- [boson] (a3),
				(v5) -- [boson] (a4),
				(v5) -- [boson] (a5),
				(v6) -- [boson] (a6),
				(v6) -- [boson] (a7),
				(v6) -- [boson] (a8),
				(v7) -- [boson] (a9),
				(v7) -- [boson] (a10),
				(v7) -- [boson] (a11),
			};
			\node[rotate=25] at (-1.2,1.45) {$\cdots$};
			\node[rotate=-25] at (1.2,1.45) {$\cdots$};
			\node at (0.15,-1.7) {$\cdots$};
			\node at (0,1.3) {$\cdots$};
		\end{feynman}
	\end{tikzpicture}
	\caption{Loop corrections to the scalar field propagator in linearized gravity.}
	\label{fig:diagrams-scalar-field-propagator}
\end{figure}

The one-loop effective action contains quantum corrections which "dress" the propagator and make the couplings run. From the diagram perspective, the loops contributing to the scalar field propagator  in linearized gravity are shown in fig.\ \ref{fig:diagrams-scalar-field-propagator}.

The first diagram, label (a), is at zeroth-order in $h_{\mu\nu}$. It is quadratically divergent by power counting and renormalized by a mass term $\sqrt{g}\delta m^2 \varphi^2$.
The $\sqrt{g}$ is needed for covariance of the counterterm. The expansion of $\sqrt{g}$ in the action generates the family (c), where the diagram (a) represents the zero-order.
Due to the absence of external momentum running in the loop, the renormalization of the diagrams (a) and (c) just shifts the value of the effective mass but induces no running.

The situation is different for diagrams (b), where $h_{\mu\nu}$ tails inject external momentum in the loop. According to power counting, diagram (b) is quadratically divergent because the gravity vertex is quadratic in derivatives. The counterterm must be $\sqrt{g}\delta\xi R \varphi^2$ for dimensional and covariant reasons. The Ricci scalar $R$ is the unique curvature that is a scalar and has the appropriate dimension of 2.
Also, $R$ contains derivatives of the metric which inject the external, metric momentum into the loop, unlike to $\sqrt{g}$ which has no derivatives. 
This is why the mass counterterm is not adequate to renormalize diagrams (b).

Other diagrams generated by the metric expansion $g_{\mu\nu}=\delta_{\mu\nu}+h_{\mu\nu}$ are shown in the second line of fig.\ \ref{fig:diagrams-scalar-field-propagator}.
Diagrams (d) are complementary to diagrams (b), thus they are also renormalized by the counterterm $\sqrt{g}\delta\xi R \varphi^2$.
Diagrams (e) and (f) are similar to diagram (b). Each insertion of an $h_{\mu\nu}$ vertex generates a new propagator in the loop, and its vertex carries a quadratic momentum dependence, just like the propagator. Therefore, these diagrams are also quadratically divergent, containing logarithmic divergences as well.
They are still renormalized by $\sqrt{g}\delta \xi R\ph^2$ counterterm.

	Diagrams (e) and (f) introduce additional structures in the finite part of the effective action compared with diagram (b). For example, family (f), with $n$ gravitational vertices, generates contributions schematically of the form $\sqrt{g}\,\xi^n\varphi^2\,F_n(\Box_1,\ldots,\Box_n)\,R_1\cdots R_n$, together with terms of lower order in curvature. Here, $F_n$ denotes a form factor whose operator arguments $\Box_i$ act on the corresponding curvature factors $R_i$, following the notation used in the covariant curvature expansion of ref.~\cite{Barvinsky:2009CPT}.

\begin{figure}[H]
	\centering
	\begin{tikzpicture}
		\begin{feynman}
			\vertex (v1) at (0,-0.9);
			\vertex (v2) at (-1,0);
			\vertex (v3) at (1,0);
			\vertex (v4) at (0,0.9);
			\vertex (a1) at (-2,-0.9);
			\vertex (a2) at (2,-0.9);
			\vertex (a3) at (0,1.9);
			
			\diagram* {
				(a1) -- [plain] (v1),
				(v1) -- [plain] (a2),
				(v2) -- [half left](v3),
				(v2) -- [half right] (v3),
				(v4) -- [boson] (a3),
			};
		\end{feynman}
	\end{tikzpicture}
	\caption{$\cl{O}(h)$ diagram contributing to the scalar field propagator.}
	\label{fig:bomb-diagram}
\end{figure}
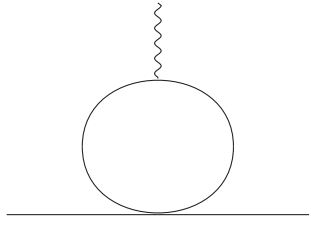

We compute explicitly the first-order diagram of family (b), shown in fig.\ \ref{fig:bomb-diagram}. 
This diagram is representative of the one-loop quantum corrections to $\xi$ (at first order in $\xi$), as we show in the following sections.
Divergences are regularized via dimensional regularization, while the finite part is investigated in the UV and IR limits to verify the decoupling process. 

\section{Tadpole diagram in linearized gravity}\label{sec:3}
Diagrams depicted in fig.\ \ref{fig:diagrams-scalar-field-propagator} are constructed from combinations of vertices and propagators derived from the action \eqref{eq:action-Euclidean}. 
In the path integral approach, they are generated from the expansion of $\int \mathcal{D}\ph e^{-\cl{S}[\ph,g]}\ph \ph$ in $\lambda$ and $h_{\mu\nu}$.
Let us clarify that the expansion in the coupling $\lambda$ is the usual perturbative expansion for small couplings adopted in QFT, while the expansion in $h_{\mu\nu}$ is the core of the linearized gravity approach. Indeed, the diagrams in fig.\ \ref{fig:diagrams-scalar-field-propagator} are at one loop and up to infinitely many orders in $h_{\mu\nu}$.

Propagators and vertices in curved spacetime are defined analogously to those in flat spacetime \cite{BirrellDavies}.
The propagator is the inverse of the Hessian operator, $H(x)$, derived from the action. For the action \eqref{eq:action-Euclidean} it is
\begin{align} \label{eq:Hessian-scalar-field}
		& \hat{H}(x,y) = \eval{\frac{\delta^{2}\mathcal{S}}{\delta\ph(x)\,\delta\ph(y)}}_{\ph=0} = H(x) \delta^4(x-y) \notag \\
		& \text{with} \qquad H(x) = \sqrt{g}\qty(-\square +m^2 +\xi R) \,.
\end{align}
The propagator $G(x,y)$ is the solution of $H(x)G(x,y)=\delta(x-y)$. 

For evaluating the loop in fig.\ \ref{fig:bomb-diagram} it is sufficient to solve the propagator equation at $\cl{O}(h)$. Alternatively, the same result can be obtained by expanding the action \eqref{eq:action-Euclidean} from the beginning.
The expansion of curvatures for $g_{\mu\nu}=\delta_{\mu\nu}+h_{\mu\nu}$ is the core of the linearized metric approach, thus we report the expansion of useful quantities in Appendix \ref{app:A1}. The zero and first order of the Hessian read, respectively,
\begin{align} 
		H_0(x) =& -\partial^2 +m^2 \,, \label{eq:Hessian-expansion}\\
		H_1(x) =& \frac{1}{2}h(-\partial^2 +m^2) +\Big( h^{\mu\nu}\partial_\mu\partial_\nu +(\partial_\mu h^{\mu\nu})\partial_\nu -\frac{1}{2}(\partial^\mu h)\partial_\mu \Big) \nn\\
		& +\xi \left(\partial_\mu\partial_\nu h^{\mu\nu} -\partial^2 h\right) \,. \label{eq:Hessian-expansion-H1}
\end{align}
The propagator is obtained perturbatively as $G=G_0+G_1$.

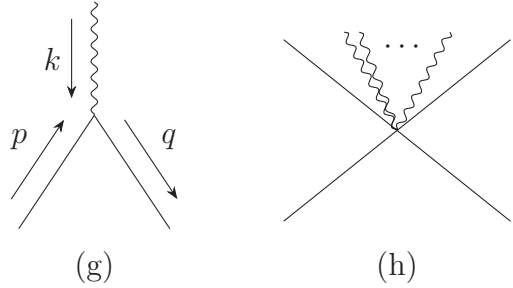
\begin{figure}
		\centering
		\begin{tikzpicture}
			\begin{feynman}
				\vertex (a1) at (0,0);
				\vertex (a2) at (2,0);
				\vertex [dot] (v1) at (1,1.5);
				\vertex (out1) at (1,3);
				\node (name) at (1,-0.5) {(g)};
				
				\diagram* {
					(a1) -- [plain, momentum=\(p\)] (v1),
					(v1) -- [plain, momentum=\(q\)] (a2),
					(v1) -- [boson, reversed momentum=\(k\)] (out1),
				};
			\end{feynman}
		\end{tikzpicture}
		\hspace{1cm}
		\begin{tikzpicture}
			\begin{feynman}
				\vertex (a1) at (-1.5,1.2);
				\vertex (a2) at (-1.5,-1.2);
				\vertex (a3) at (1.5,1.2);
				\vertex (a4) at (1.5,-1.2);
				\vertex (a5) at (-0.7,1.3);
				\vertex (a6) at (-0.5,1.3);
				\vertex (a7) at (0.7,1.3);
				\vertex (v1) at (0,0);
				\node at (0.1,1.1) {$\cdots$};
				\node (name) at (0,-1.8) {(h)};
				
				\diagram* {
					(a1) -- (v1),
					(a2) -- (v1),
					(v1) -- (a3),
					(v1) -- (a4),
					(v1) -- [boson](a5),
					(v1) -- [boson](a6),
					(v1) -- [boson](a7),
					
				};
			\end{feynman}
		\end{tikzpicture}

	\caption{(g) First-order propagator. (h) Vertex $\sqrt{-g}\lambda \ph^4$ with metric expansion.
	}
		\label{fig:first-order-propagator}
	\end{figure}

The zero-order equation is $H_0G_0 = \delta$ and can be solved in momentum space,
\begin{align} \label{eq:scalar-field-propagator-flat-spacetime}
	G_0(x,y) = \int \frac{d^4p}{(2\pi)^4}e^{i \delta^{\mu\nu}p_\mu(x_\nu-y_\nu)}\tilde{G}_0(p) \,,
	\qquad \tilde{G}_0(p) =  \frac{1}{p^2+m^2} \,.
\end{align}
The sign convention we adopt uses a positive sign in the exponential. 
The propagator at first-order is instead
\begin{align} \label{eq:first-order-scala-propagator-coordinate-space}
	G_1(x,y) \,=\, - \int d^4z\, G_0(x,z) H_1(z) G_0(z,y) \,.
\end{align}
The dependence on the external metric perturbation $h_{\mu\nu}$ enters into $H_1(z)$. We use a Fourier decomposition for $h_{\mu\nu}$,
\begin{align} \label{eq:metric-perturbation-Fourier-representation}
	& h^{\mu\nu}(z) \,=\, \int\frac{d^4k}{(2\pi)^4}e^{-ik.z}\tilde{h}^{\mu\nu}(k) \,,
\end{align}
where $.$ denotes a scalar product and $k$ is an external momentum contributing to the scalar field propagator $G_1(x,y)$. In our sign convention the momentum $k^\mu$ enters the propagator.
In momentum space, the first-order propagator \eqref{eq:first-order-scala-propagator-coordinate-space} reads
\begin{align}\label{eq:first-order-propagator-momentum-space}
	&G_1(x,y) =-\,\iint \frac{d^4p\,d^4k}{(2\pi)^8}\frac{e^{ip.x}e^{-i(p+k).y}}{(p^2+m^2)[(p+k)^2+m^2]}\tilde{h}^{\mu\nu}(k)V_{\mu\nu}(p,k)
\end{align}
where
\begin{align} \label{eq:vertex-from-Hessian-momentum-space}
	V_{\mu\nu}(p,k) 
		=\frac12 \delta_{\mu\nu} (p^2+p.k+m^2) -p_\mu (p_\nu +k_\nu) - \xi(k_\mu k_\nu - \delta_{\mu\nu}k^2) \,.
\end{align}
The tensor structure $V_{\mu\nu}$ represents the first-order vertex shown in \ref{fig:first-order-propagator} (g). It follows from the momentum representation of \eqref{eq:Hessian-expansion-H1} in the sign convention \eqref{eq:scalar-field-propagator-flat-spacetime}.
The momentum $q$ in the figure is rewritten by momentum conservation.
The same picture is also the diagrammatic representation of $G_1$.
	Finally, the quartic vertex $\sqrt{g}\lambda \ph^4/4!$ has a metric expansion depicted in fig.\ \ref{fig:first-order-propagator} (h). For the purpose of computing the diagram in fig.\ \ref{fig:bomb-diagram}, we need the zeroth order in $h_{\mu\nu}$, i.e.\ the flat spacetime version $\lambda \phi^4/4!$.

We can now build the diagram of our interest,
%
\begin{align}
	D(x,y)= - N(\la) \int \dd[4]{z_2} G_0(x,z_2) G_1(z_2,z_2) G_0(z_2,y) \qquad \tx{with} \qquad N(\lambda)\equiv \frac{\la}{2} \,.
\end{align}
Using the momentum representation for the propagators,
\begin{align}
	D(x,y)= & N(\la) \int\frac{\dd[4]{p_1}}{(2\pi)^4} e^{ip_1.x}\tilde{G}_0(p_1) e^{-i(p_1+k).y} \tilde{G}_0(p_1+k) \nn \\
	& \times \iint \frac{d^4p\,d^4k}{(2\pi)^8} \tilde{G}_0(p) \tilde{h}^{\mu\nu}(k)V_{\mu\nu}(p,k) \tilde{G}_0(p+k)  \,,
\end{align}
where $p_1$ and $k$ are the external momenta, and $V_{\mu\nu}$ is the vertex structure \eqref{eq:vertex-from-Hessian-momentum-space}.

	Let us now focus on the loop structure,
	\begin{align}\label{eq:loop-bomb-diagram-momentum-space}
		I_{\mu\nu}(k) =N(\la)\int\frac{d^4p}{(2\pi)^4}\frac{1}{p^2+m^2}V_{\mu\nu}(p,k)\frac{1}{(p+k)^2+m^2} \,.
	\end{align}
	By power counting, \eqref{eq:loop-bomb-diagram-momentum-space} has divergences up to quadratic order. We will evaluate it explicitly using dimensional regularization, where polynomial divergences are suppressed and logarithmic ones are parametrized as poles $\ep=(n-4)$, with $n$ the dimensionality parameter. We will comment on the divergent and finite part of the loop.
	In \eqref{eq:loop-bomb-diagram-momentum-space}, $p$ is the internal momentum and $k$ is the momentum of the external $h_{\mu\nu}$ which, in our sign convention, enters the loop. 
	
	Let us stress that the quantum correction \eqref{eq:loop-bomb-diagram-momentum-space} is non-covariant since it is based on an explicit split of the metric, which is not guaranteed to hold in any reference frame. Even so, eq.\ \eqref{eq:loop-bomb-diagram-momentum-space} provides guidance for extrapolating the covariant, one-loop effective action. 
	Since \eqref{eq:loop-bomb-diagram-momentum-space} is a correction to the two-point function, it provides information about the effective action quadratic in $\varphi$. 
	The coefficient needed to map the loop to the effective action follows from the propagator equation. 
	The free propagator $G_0(x,y)$ is derived from the action \eqref{eq:action-Euclidean} with the convention of having a $1/2$ prefactor for the kinetic term.
	The first-order propagator $G_1(x,y)$ is obtained expanding $(H_0+H_1)^{-1}\approx H_0^{-1} - H_0^{-1}H_1 H_0^{-1}$, thus its contribution to the effective action is mapped multiplying by $-1/2$. 
	
\subsection{The divergent Effective Action from the tadpole}
The integral \eqref{eq:loop-bomb-diagram-momentum-space} can be unpacked into easier integrals that are standard knowledge in QFT (see e.g. \cite{BuchbinderShapiro:2021} and \cite{Peskin1995}). We give full details on how to compute \eqref{eq:loop-bomb-diagram-momentum-space} in Appendix \ref{app:A1}. There, we use dimensional regularization moving from dimension 4 to $2\om$ with $\om$ a complex number to parameterize divergences with analytic continuation. The divergent part of \eqref{eq:loop-bomb-diagram-momentum-space} was derived in eq.\ \eqref{eq:divergent-part-diagram} and we recall it below, 
	\begin{align} \label{eq:bomb-diagram-divergent-structure-text}
		I_{\mu\nu}^{\text{div}}(k) = \frac{N(\la)}{\varepsilon} \qty(\xi-\frac16) (\delta_{\mu\nu} k^2-k_\mu k_\nu) \,,
	\end{align}
where $\varepsilon$ is a pole that diverges as $\om\to 2$, as defined in \eqref{eq:pole-definition-varepsilon}.
	
	Let us discuss whether the divergent structure \eqref{eq:bomb-diagram-divergent-structure-text} is in agreement with the counterterm elaborated in the Introduction for diagrams (b) and (c) in fig.\  \ref{fig:diagrams-scalar-field-propagator}.
	Firstly, the tensor structure is the momentum representation of the first-order expansion of $R$ in flat-spacetime, as can be seen in \eqref{eq:expansion-Ricci-scalar-flat-background}. This suggests that the divergent effective action, obtained summing up the Feynman diagrams at all orders in $h_{\mu\nu}$, will contain $\sqrt{g}R$. Secondly, we found the conformal factor $\xi - 1/6$, which is expected in order to preserve conformal symmetry of UV divergences \cite{Buchbinder:1985ew}.
	Since $\xi - 1/6$ is not an output of the curvature expansion \eqref{eq:expansion-Ricci-scalar-flat-background}, it is reasonable to believe that such a prefactor will be present in any divergent part of diagrams \ref{fig:diagrams-scalar-field-propagator}.
	More generally, the whole prefactor of \eqref{eq:expansion-Ricci-scalar-flat-background}, up to the first-order expansion of $\sqrt{g}R$, will be preserved at any order in $h_{\mu\nu}$. Thus we obtained the quantum output of the prefactor from the first order only. 
	
	The one-loop divergent effective action suggested by the loop divergence \eqref{eq:bomb-diagram-divergent-structure-text} is thus
	\begin{align} \label{eq:divergent-one-loop-EA}
		\bar{\Ga}^{(1)}_{\text{div}} = -\frac{1}{2(4\pi)^2(2-\om)} \int \dd[4]{x}\sqrt{g} \,\ \frac{\la}{2}\qty(\xi-\frac16)R\varphi^2 \,,
	\end{align}
	where we substituted the pole with \eqref{eq:pole-definition-varepsilon} and took the limit $\omega\to 2$ everywhere possible. 
	An overall factor $-1/2$ was introduced for mapping the loop correction to the effective action, in accordance with the convention used in the action \eqref{eq:action-Euclidean}.

	This result is in perfect agreement with the one-loop divergent effective action derived in \cite{deBerredo-Peixoto:2003pkj} using the covariant heat-kernel technique.
	
	Although the loop \eqref{eq:bomb-diagram-divergent-structure-text} was derived in non-covariant linearized gravity, the effective action \eqref{eq:divergent-one-loop-EA} extracted from that was formulated by requiring covariance and locality.
	The agreement between \eqref{eq:divergent-one-loop-EA} and \cite{deBerredo-Peixoto:2003pkj} is a double check for the heat-kernel technique and the by-hand covariantization of loop corrections computed in the linerized gravity approach.
	
	From the divergences one can start the renormalization procedure. Thus, the general counterterm needed to renormalize the diagrams in fig.\ \ref{fig:diagrams-scalar-field-propagator} is given by $\sqrt{g}\delta \xi R\ph^2$.

\subsection{The finite Effective Action from the tadpole}
	The explicit computation of the finite part of the loop is carried out in Appendix \ref{App:finite-part}. Here we report the final result \eqref{eq:finite-part-loop} in momentum space and in the limit $\omega\to 2$,
	\begin{align} \label{eq:finite-loop-4dim}
		I_{\mu\nu}^{\text{fin}}(k) = \frac{N(\la)}{(4\pi)^2} \qty[\qty(\xi-\frac16)\qty(\ga+\ln\qty(\frac{m^2}{4\pi\mu^2})) +\xi F_1(a) + F_2(a)] \qty[k_\mu k_\nu -\delta_{\mu\nu} k^2] \,,
	\end{align}
	where the form factors were computed in \eqref{eq:F1+F2-form-factors},
	\begin{align} \label{eq:form-factor-finite-part}
		\xi F_1(a)+F_2(a) = -\frac{1}{18} -\qty(2\xi -\frac12 +\frac{2}{3a^2}) Y(a) \,, \qquad Y \equiv 1-\frac{1}{a}\ln\qty(\frac{2+a}{2-a}) \,.
	\end{align}
	Here, $a^2\equiv 4k^2/(4m^2+k^2)$ was introduced in \eqref{eq:Y-a-definition}, and $\gamma$ is the Euler-Mascheroni constant. 
	The first term in brackets in eq.\ \eqref{eq:finite-loop-4dim} is independent of the external momentum. Hence, it does not contribute to the beta function in the momentum-subtraction scheme. We therefore omit it only in the analysis of the momentum-dependent form factor, while retaining it when comparing with the effective potential.
	
	Starting from \eqref{eq:finite-loop-4dim} and using covariance of the action again, we can deduce the finite part of the one-loop effective action for the scalar field to be
	\begin{align} \label{eq:effective-action-finite-part}
		\bar{\Ga}_{\text{fin}}^{(1)} = \frac{1}{2(4\pi)^2} \int\dd[4]{x}\sqrt{g} \,\ \la \varphi^2\Big[ \frac12\qty(\xi-\frac16)&\qty(\ga+\ln\qty(\frac{m^2}{4\pi\mu^2})) -\frac{1}{36} \nn \\
		&-\qty(\xi-\frac16)Y +\frac{(a^2-4)}{12a^2}Y \Big] R  \,.
	\end{align}
	We manipulated the form factor \eqref{eq:form-factor-finite-part} in order to allow a direct comparison with the effective action of \cite{deBerredo-Peixoto:2003pkj}.
The term $\sqrt{g}R$ is inferred from the first-order expansion of $R$ appearing in \eqref{eq:form-factor-finite-part}. An overall factor $-1/2$ multiplies the loop as discussed for the divergent part.
The action \eqref{eq:effective-action-finite-part} is in agreement with \cite{deBerredo-Peixoto:2003pkj}.
	In \eqref{eq:effective-action-finite-part}, the form factors are written in coordinate space,
	\begin{align}
		a^2= \frac{-4\square}{4m^2-\square} \,,
	\end{align}
where $\square$ is the d'Alembert operator.
As for the divergent part of the effective action, we extrapolated the finite part by computing just one diagram.

Let us comment that the calculation in \cite{deBerredo-Peixoto:2003pkj} is up to second-order in the field strenghts (that we call curvatures), i.e.\ $R_{\mu\nu\alpha\beta}$ and $(1/6-\xi)R + \la \ph^2/2$. It is based on the covariant expansion of the heat-kernel at second order in curvatures elaborated in \cite{Barvinsky:1990up}. 
Therefore, any comparison we can make between our diagrammatic approach and ref.\ \cite{deBerredo-Peixoto:2003pkj} is up to second order in curvatures.
This means that contributions from diagrams (e) and (f) in fig.\ \ref{fig:diagrams-scalar-field-propagator} are not necessary for the sake of comparison with \cite{deBerredo-Peixoto:2003pkj}.

\subsection{UV, decoupling and running}

	We are now in a position to study the ultraviolet and infrared limits of the $\xi$-sector of the effective action. Equivalently, these limits can be analyzed at the level of the loop, exploiting the simpler framework provided by momentum space.

The UV limit is achieved for $k^2\gg m^2$. Using the dimensionless variable $\tau=k^2/m^2$, we expand the form factor \eqref{eq:form-factor-finite-part}. We have $ a \approx 2(1-2/\tau)$ and $Y \approx 1-(1/2)\ln\tau$. Therefore,
\begin{align}
\xi F_1(a)+F_2(a) \approx \qty(\frac{5}{18}-2\xi) +\qty(\xi-\frac16) \ln\tau \,.
\end{align}
In the UV, the logarithm $\ln(\tau)$ combines with the logarithm $\ln(m^2/\mu^2)$, establishing a relation between the UV scale $k^2$ and the arbitrary scale $\mu^2$ introduced by the regularization. Moreover, the structure $\xi-1/6$ is preserved, highlighting that the fields behave as effectively massless at high energies.

Conversely, the IR limit corresponds to $k^2 \ll m^2$. In this regime, $a^2 \approx \tau$ and $Y \approx -\tau/12-\tau^2/80$. The form factors then reduce to
\begin{align}
\xi F_1(a)+F_2(a) \approx \qty(\xi -\frac15) \frac{\tau}{6} \,.
\end{align}
It is worth noting that the zeroth-order term, $\mathcal{O}(\tau^0)$, cancels out, leading to the expected decoupling behavior described by the Appelquist--Carazzone theorem \cite{Appelquist:1974tg}.

Another important aspect that can be studied at this stage of the calculation is the running of $\xi$. To capture the physical behavior at different energy scales, we adopt the momentum-subtraction (MOM) scheme. In contrast to mass-independent schemes such as the minimal subtraction ($\overline{\mathrm{MS}}$), where the renormalization-group evolution is conventionally expressed in terms of the arbitrary renormalization scale $\mu$, in the MOM scheme the renormalized coupling is defined through a subtraction condition at a physical momentum scale $k$. Accordingly, the renormalization-group equation can be parametrized in terms of the subtraction scale $k$ and written as
\begin{align} \label{eq:renormalization-group-equation}
\left(k^2 \frac{\pa}{\pa k^2} + \be_\xi^{\text{MOM}} \frac{\pa}{\pa \xi} +\cdots\right) \,  \Ga_R = 0\,,
\end{align}
where $\Ga_R = \mathcal{S} + \bar\Ga^{(1)}_{\tx{fin, } R}$ is the renormalized effective action, and $\be_\xi^{\text{MOM}} = k^2 \pdv{\xi}{k^2}$.

	The dots in \eqref{eq:renormalization-group-equation} stand for the running of the other theory's parameters. At first order in $\la$, the beta functions $\be_{m^2}^{\text{MOM}}$ and $\be_\la^{\text{MOM}}$ can be safely decoupled from the evaluation of $\be_\xi^{\text{MOM}}$. 
	First, the running of the quartic coupling is induced by diagrams of $\mathcal{O}(\la^2)$, which lie beyond the linear approximation adopted here. Moreover, the derivative with respect to $\la$ generates $\ph^4$ terms that do not mix with the $R\ph^2$ structure.
Second, the derivative with respect to the mass, $\be_{m^2}^{\text{MOM}} \frac{\pa}{\pa m^2}$, generates terms proportional to $\sqrt{g}\ph^2$. Since this geometric structure is fundamentally different from the curvature-dependent non-minimal sector $\sqrt{g}R\ph^2$, these terms also do not mix at one-loop level. Therefore, the running of both $\la$ and $m^2$ can be safely discarded, and the running of $\xi$ is driven purely by the explicit momentum dependence encoded in the non-local form factors of the non-minimal term $\xi R\ph^2$.

Consequently, we obtain the relation
\begin{align}
k^2 \frac{\pa}{\pa k^2} \bar\Ga^{(1)}_{\text{fin, }R} + \be_\xi^{\text{MOM}} \frac{\pa}{\pa \xi}\mathcal{S} = 0 \,.
\end{align}
Let us comment that the term $\pdv*{\bar{\Gamma}^{(1)}_{\tx{fin, }R}}{\xi}$ induces an $\cl{O}(\lambda^2)$ contribution to the beta function $\beta_\xi$ and is thus neglected. 

From the classical action, the variation with respect to the coupling yields
\begin{align}\label{beta_class}
\be_\xi^{\text{MOM}} \frac{\pa}{\pa \xi}\mathcal{S} &=  \be_\xi^{\text{MOM}} \frac{\pa}{\pa \xi}\int d^4x \sqrt{g} \left[\frac{1}{2}\xi R\ph^2\right]= \int d^4x \sqrt{g} \ph^2\left[\frac{1}{2}\be_\xi^{\text{MOM}} \right]R\,.
\end{align}

On the other hand, the renormalized, finite effective action $\bar\Ga^{(1)}_{\tx{fin, } R}$ in the MOM scheme is obtained by subtracting from the regularized effective action \eqref{eq:effective-action-finite-part} at a scale $p$ the same quantity at a different scale $k$. 
The relevant part that remains from such a procedure is contained in the difference between the form factors \eqref{eq:form-factor-finite-part} evaluated at the two scales,
\begin{equation}
\mathcal{F}(p,k) \equiv \xi F_1(a(p)) + F_2(a(p)) - \xi F_1(a(k)) - F_2(a(k)) \,.
\end{equation}
Taking the derivative with respect to the physical momentum scale $k$, we find
\begin{align}
k^2 \frac{\pa }{\pa k^2}\mathcal{F}(p,k) &= k^2 \frac{\pa }{\pa k^2}\left\{\frac{\la}{4(4\pi)^2}\int_0^1 d\al (\xi - \al(1-\al))\log\left[\frac{\al(1-\al)p^2 + m^2}{\al(1-\al)k^2 + m^2}\right]\right\}\nonumber\\
&=-\frac{\la}{4(4\pi)^2}\int_0^1 d\al [\xi - \al(1-\al)]\frac{\al(1-\al)}{\al(1-\al)+\frac{m^2}{k^2}}\,.
\end{align}
Therefore, the beta function for the non-minimal coupling $\xi$ in the MOM scheme is given by
\begin{align}
\be_\xi^{\text{MOM}}(k^2) &= \frac{\la}{2(4\pi)^2}\int_0^1 d\al [\xi - \al(1-\al)]\frac{\al(1-\al)}{\al(1-\al)+\frac{m^2}{k^2}}\,.
\end{align}
We can now evaluate this exact one-loop beta function at the external momentum scale $p^2$. It corresponds to taking the subtraction scale $k^2$ to be equal to the external momentum scale, thus promoting $k^2$ to be a physical scale. We obtain
\begin{align}
\be_\xi^{\text{MOM}}(p^2)
=
\frac{\la}{32\pi^2}
\int_0^1 d\al\,
\frac{
\al(1-\al)\left[\xi-\al(1-\al)\right]
}{
\al(1-\al)+\frac{m^2}{p^2}
}\,.
\end{align}

We can now analyze the asymptotic behavior of this expression. In the ultraviolet (UV) limit, where $p^2\gg m^2$, the mass term in the denominator is suppressed, yielding
\begin{align}
\be_\xi^{\text{UV}}
&\approx
\frac{\la}{32\pi^2}
\int_0^1 d\al\,
\left[\xi-\al(1-\al)\right]
\nonumber\\
&=
\frac{\la}{32\pi^2}
\left(\xi-\frac{1}{6}\right)\,.
\end{align}
This result is equivalent to the standard asymptotic $\overline{\mathrm{MS}}$ beta function obtained in Ref.~\cite{deBerredo-Peixoto:2003pkj}, after accounting for the different conventions used for the logarithmic derivative. Indeed,
\begin{align}
p^2\frac{d}{dp^2}
=
\frac{1}{2}p\frac{d}{dp}\,,
\end{align}
and therefore
\begin{align}
\be_\xi^{\overline{\mathrm{MS}}}
=
2\be_\xi^{\text{UV}}
=
\frac{\la}{16\pi^2}
\left(\xi-\frac{1}{6}\right)\,.
\end{align}

Conversely, in the infrared (IR) limit, where $p^2\ll m^2$, we expand the denominator in powers of $p^2/m^2$, obtaining
\begin{align}
\be_\xi^{\text{IR}}
&\approx
\frac{\la}{32\pi^2}
\frac{p^2}{m^2}
\int_0^1 d\al\,
\al(1-\al)
\left[\xi-\al(1-\al)\right]
\nonumber\\
&=
\frac{\la}{32\pi^2}
\left[
\frac{1}{6}
\left(\xi-\frac{1}{5}\right)
\right]
\frac{p^2}{m^2}
+
\mathcal{O}\left(\frac{p^4}{m^4}\right).
\end{align}
This result makes the Appelquist--Carazzone decoupling explicit: the contribution of the massive scalar fluctuations to the running of $\xi$ vanishes quadratically with the external momentum in the low-energy limit. 

After accounting for the normalization convention of the logarithmic derivative discussed above, these results agree with \cite{deBerredo-Peixoto:2003pkj}. In that reference, the physical renormalization-group scheme is referred to as mass-dependent, emphasizing that the beta function depends explicitly on the ratio $p^2/m^2$.

\section{Effective Potential in linearized gravity}\label{sec:4}
	As a cross-check of our previous calculations, and to further demonstrate the versatility of the linearized gravity approach, we now compute the effective potential for the scalar field action \eqref{eq:action-Euclidean}. Let us start by recalling what the effective potential is. 
	The one-loop effective action derived from \eqref{eq:action-Euclidean}, within the background field method and the path integral approach, is
	\begin{align} \label{eq:EA-1loop-scalar-field}
		\Ga[\ph] = S[\ph] + \frac12 \Tr \log\left(\frac{\delta^2 S[\Phi]}{\delta \Phi \delta \Phi}\right)\Bigg|_{\Phi=\ph}\,,
	\end{align}
	where $\ph$ is the classical, background field \cite{BuchbinderShapiro:2021}.
	The effective potential is defined as the purely non-derivative part of the effective action (incorporating both the classical potential and its quantum corrections), thus excluding the kinetic and higher-order derivative terms. That is
	\begin{align} \label{eq:effective-potential-formal-def}
		V_{eff} = \int \dd[4]{x} \sqrt{g} \,\mathcal{V}_{eff}\,=\, \Ga[\ph] \Big|_{\ph \, = \, \text{const}} \,.
	\end{align}
	In this case the classical action evaluated at a constant, background field is given by
	\begin{align}
		S[\ph] \Big|_{\ph \, = \, \text{const}}  = \int \dd[4]{x}\sqrt{g}  \left\{\frac{1}{2}m^2\ph^2+\frac{1}{2}\xi R\ph^2 +\frac{\lambda}{4!}\ph^4 \right\}
	\end{align}
	and the bilinear form is
	\begin{align}\label{hessian}
		H\,=\,\frac{\delta^2 S[\Phi]}{\delta \Phi \delta \Phi}\Bigg|_{\Phi=\ph = \text{const}} = \sqrt{g}\left[-\Box+M^2+\xi R\right]\,, 
		\qquad \text{with} \qquad 
		M^2 = m^2 + \frac{\la}{2}\ph^2 \,.
	\end{align}
	We collected all constant terms into a shifted mass $M$. 
	Our goal now is to explicitly compute the effective potential \eqref{eq:effective-potential-formal-def} employing the linearized gravity approach, as done for deriving the loop corrections. 
	Although the calculations will be similar to the previous ones, since the initial action \eqref{eq:EA-1loop-scalar-field} is the same, in the effective potential case we have to evaluate the trace-logarithm of the Hessian \eqref{hessian} which requires some attention. 
	
	By using the linearized gravity approach we start by expanding quantities in orders of $h_{\mu\nu}$ and move to momentum space for elaborating the equations. The final result will be a perturbative effective potential
	\begin{align}
		V_{eff} = V_{eff}^{(0)} + V_{eff}^{(1)} + V_{eff}^{(2)} + \cdots
	\end{align}
	The index $^{(i)}$ represents the $i$th order in $h$.
The first step is the expansion of the Hessian. The structure of $H_0$ and $H_1$ was already reported in Section \ref{sec:3} and it holds here with the attention of substituting $m^2$ with $M^2$. Let us now add the second order, 
	\begin{align}
		H_2 \, = \, \left(-\frac{1}{4}h_{\mu\nu}^2 + \frac{1}{8}h^2\right)\left(-\pa^2+M^2\right) + \frac{1}{2}h\left(-\Box^{(1)}+ \xi R^{(1)}\right)+ \left(-\Box^{(2)}+ \xi R^{(2)}\right)\,.
	\end{align}
The explicit expression for the d'Alambert and $R$ can be found in Appendix \ref{app:A1}.

The key idea for computing the effective potential in linearized gravity is to also expand the logarithm in \eqref{eq:EA-1loop-scalar-field} in orders of $h_{\mu\nu}$,
	\begin{align} \label{eq:TrLog-expansion}
		\frac12 \Tr\log( H_0 +H_1+H_2 ) &= \frac12 \Tr\log H_0 +\frac12 \Tr\log[ 1 + G_0 \qty( H_1+H_2)] \notag \\
		&= \frac12 \Tr \log H_0 +\frac12 \Tr (G_0H_1) \notag \\
		&\quad +\frac12 \Tr(G_0 H_2) -\frac14 \Tr( G_0H_1G_0H_1 ) \,,
	\end{align}
	where we introduced $G_0=H_0^{-1}$ which depends on the shifted mass $M$. We now evaluate \eqref{eq:TrLog-expansion} order by order in dimensional regularization.
	
	For the zero order, we calculate the following integral\footnote{The functional trace definition is $\Tr A(x,y) = \int \dd[4]{x} \lim_{y\to x} A(x,y)$.}
	\begin{align}
		\frac12 \Tr \log \left(-\partial^2 + M^2\,\right) 
		\, &=\, \frac12 \int \dd[4]{x} \int\frac{d^{2\om}p}{(2\pi)^{2\om}}\log\left(p^2+M^2\right)\,, \nn \\
		\, &=\, \frac12 \int \dd[4]{x} \frac{M^4}{2(4\pi)^2}(\mu^2)^{\om-2} \left[ -\frac{1}{2-\omega} - \frac{3}{2} + \gamma + \ln\left( \frac{M^2}{4\pi\mu^2} \right) \right] \,.
	\end{align}
The details of the calculation are in Appendix \ref{App:eff-pot}. We can assign the diagrammatic representation of fig.\ \ref{fig:effective-potential-orders} (a) to the previous integral. 
The zeroth-order effective potential is then given by
	\begin{align}\label{0potential}
		V_{eff}^{(0)}\,&=\, S^{(0)}[\ph]+\frac12 \Tr \log\left(-\partial^2 + M^2\,\right) \nn\\
		&=\int d^4x  \left\{\frac{1}{2}m^2\ph^2+\frac{\lambda}{4!}\ph^4 + \frac{M^4}{2(4\pi)^2}(\mu^2)^{\om-2} \left[ -\frac{1}{2-\omega} - \frac{3}{2} + \ga+\ln\left(\frac{M^2}{4\pi\mu^2}\right) \right]\right\}\,.
	\end{align}
	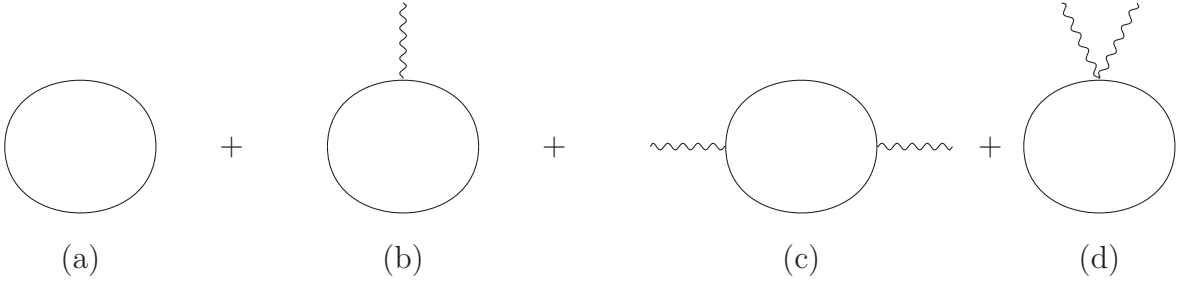
\begin{figure}
		\centering
		\begin{tikzpicture}
			\begin{feynman}
				\vertex (v1) at (0,-0.9);
				\vertex (v2) at (-1,0);
				\vertex (v3) at (1,0);
				\vertex (v4) at (0,0.9);
				\node (name) at (0,-1.5) {(a)};
				\node (plus) at (2,0) {+};
				
				\diagram* {
					(v2) -- [half left](v3),
					(v2) -- [half right] (v3),
				};
			\end{feynman}
		\end{tikzpicture}
		\qquad
		\begin{tikzpicture}
			\begin{feynman}
				\vertex (v1) at (0,-0.9);
				\vertex (v2) at (-1,0);
				\vertex (v3) at (1,0);
				\vertex (v4) at (0,0.9);
				\vertex (a1) at (-2,-0.9);
				\vertex (a2) at (2,-0.9);
				\vertex (a3) at (0,1.9);
				\node (name) at (0,-1.5) {(b)};
			\node (plus) at (2,0) {+};
				
				\diagram* {
					(v2) -- [half left](v3),
					(v2) -- [half right] (v3),
					(v4) -- [boson] (a3),
				};
			\end{feynman}
		\end{tikzpicture}
		\qquad
		\begin{tikzpicture}
			\begin{feynman}
				\vertex (v2) at (-1,0);
				\vertex (v3) at (1,0);
				\vertex (a1) at (-2,-0.9);
				\vertex (a3) at (-2,0);
				\vertex (a4) at (2,0);
				\node (name) at (0,-1.5) {(c)};
				\node (plus) at (2.5,0) {+};
				
				\diagram* {					
					(v2) -- [half left](v3),
					(v2) -- [half right] (v3),
					(v2) -- [boson] (a3),
					(v3) -- [boson] (a4),
				};
			\end{feynman}
		\end{tikzpicture}
		\begin{tikzpicture}
			\begin{feynman}
				\vertex (v1) at (0,-0.9);
				\vertex (v2) at (-1,0);
				\vertex (v3) at (1,0);
				\vertex (v4) at (0,0.9);
				\vertex (a1) at (-2,-0.9);
				\vertex (a2) at (2,-0.9);
				\vertex (a3) at (-0.5,1.9);
				\vertex (a4) at (0.5,1.9);
				\node (name) at (0,-1.5) {(d)};
				
				\diagram* {
					(v2) -- [half left](v3),
					(v2) -- [half right] (v3),
					(v4) -- [boson] (a3),
					(v4) -- [boson] (a4),
				};
			\end{feynman}
		\end{tikzpicture}
		\caption{Vacuum bubbles for $V_{eff}$ in linearized gravity.}
		\label{fig:effective-potential-orders}
	\end{figure}

	The first-order correction in the expansion \eqref{eq:TrLog-expansion} is calculated analogously. We use the Fourier representation of $H_1,G_0$ and solve the momentum integrals. Let us stress that in the computations for the effective potential the mass $m$ is traded with $M$.
	More details are in Appendix \ref{App:eff-pot}.  Here we report the result 
	\begin{align} \label{1-correction}
		\frac12\Tr G_0H_1
		& =  \frac12 \int d^4x \Bigg\{\frac{1}{2}h(x)\frac{M^4}{2(4\pi)^2} (\mu^2)^{\omega-2} \left[ -\frac{1}{2-\omega} - \frac{3}{2} + \ga+\ln\left(\frac{M^2}{4\pi\mu^2}\right) \right]\nn\\
		& \qquad +\frac{M^2}{(4\pi)^2} (\mu^2)^{\om-2}\left[ -\frac{1}{2-\omega} - 1 +\ga+\ln\left(\frac{M^2}{4\pi\mu^2}\right) \right]\xi R^{(1)}\Bigg\} \,.
	\end{align}
	Because of the presence of $h_{\mu\nu}$, we use the diagrammatic representation fig.\ \ref{fig:effective-potential-orders} (b).
	Altogether, the first-order effective potential is given by
	\begin{align}\label{1potential}
		V_{eff}^{(1)}\,&=\, S^{(1)}[\ph]+\frac12\Tr G^{(0)}H^{(1)} \nn\\
		=& \int d^4x  \left\{\frac12h\left[\frac{1}{2}m^2\ph^2+\frac{\lambda}{4!}\ph^4\right]+\frac12\xi\ph^2R^{(1)}\right\} \nn\\
		&+\frac12 \int d^4x \Bigg\{\frac{1}{2}h\frac{M^4}{2(4\pi)^2} (\mu^2)^{\omega-2} \left[ -\frac{1}{2-\omega} - \frac{3}{2} + \ga+\ln\left(\frac{M^2}{4\pi\mu^2}\right) \right]\nn\\&+\frac{M^2}{(4\pi)^2} (\mu^2)^{\om-2}\left[ -\frac{1}{2-\omega} - 1 +\ga+\ln\left(\frac{M^2}{4\pi\mu^2}\right)\right]\xi R^{(1)}\Bigg\}\,,
	\end{align}

	Note that the factor $\xi-1/6$, which should appear in the divergent term proportional to the Ricci scalar because of the conformal symmetry of the divergences, does not appear here.
	This is not an error, and the reason is that the third line in \eqref{1potential}, where the curvature $R^{(1)}$ appears, is actually a superficial term that does not contribute to the effective potential assuming suitable boundary conditions. 
	This is a consequence of taking $\ph$ to be constant, 	since $\ph^2R^{(1)}$ is a total derivative,
	\begin{align}
		R^{(1)} = \partial_\mu(\partial_\nu h^{\mu\nu} - \partial^\mu h) \,.
	\end{align}

Moreover, the lack of an $R$-term in the first-order effective potential can be explained from a diagrammatic point of view. $V^{(1)}_{\tx{eff}}$ is represented by the diagram (b). In such diagram the external tail $h_{\mu\nu}$ does not inject any external momentum into the loop, due to momentum conservation. Therefore no $R$-term can be generated. The contribution of diagram (b) will be an $M^4$-term, as verified by the explicit calculation and shown in the second line of \eqref{1potential}.
	
	Curvature-related contributions to the effective potential are expected to appear from the second order $V^{(2)}_{\tx{eff}}$. The possible $\cl{O}(h^2)$ vacuum diagrams that contribute to $V^{(2)}_{\tx{eff}}$ are diagrams (c) and (d) of fig.\ \ref{fig:effective-potential-orders}.	
	 Again, from a diagrammatic perspective, we see that in diagram (c) the external line $h_{\mu\nu}$ does inject momentum in the loop; in diagram (d), even if it is a tadpole, the external gravity legs are not amputated and the overall diagram is not a total derivative (look at the second-order expansion of $R$ in \eqref{eq:expansion-Ricci-scalar-flat-background} in Appendix \ref{app:A1}). Therefore, diagrams (c) and (d) will furnish an $M^4$-, an $R$- and an $R^2$-contribution to the effective potential.
	
	Before proceeding to evaluate $V^{(2)}_{\tx{eff}}$ explicitly, let us give a second explanation of why the $\xi-1/6$ coefficient cannot appear in $V^{(1)}_{\tx{eff}}$ based on the scheme of calculation. 
	When we write the diagrams in momentum space and use the Feynman formula \eqref{eq:Feynman-trick-denominator} for rewriting the propagators, the numerical factor that combines with $\xi$ is generated by the $\alpha$-integral. However, the Feynman trick is used when there are two or more propagators in the loop. This is not the case for diagram (b), thus $\xi-1/6$ cannot appear.
	On the other hand, diagram (c) has two propagators and, together with its companion (d), they are expected to furnish the $\xi-1/6$ prefactor in the divergent part. Let us stress that such a prefactor appears when summing (c) and (d) together, since there are no other diagrams contributing to $V^{(2)}_{\tx{eff}}$.
	
	Let us now proceed to compute the second-order contribution to the effective potential
	\begin{align} \label{eq:Veff-2-order-impl}
		V_{eff}^{(2)} = S^{(2)}[\ph] - \frac{1}{4}\Tr G^{(0)}H^{(1)}G^{(0)}H^{(1)}+\frac12\Tr G^{(0)}H^{(2)} \,.
	\end{align}
	As already anticipated, the diagrammatic representation of the quantum corrections is shown in fig. \ref{fig:effective-potential-orders}, respectively diagrams (c) and (d). 
	
Let us stress that our purpose is to study the $\xi R\ph^2$ sector in the effective potential and compare it with the effective action. The formula \eqref{eq:Veff-2-order-impl} brings more information than the $\xi$ sector alone, for example the form factors of terms quadratic in curvature like $R f(\square) R$. In fact, it is well known that higher-order curvature terms are needed for the renormalizability of the vacuum sector \cite{BuchbinderShapiro:2021}.

In order to avoid these terms and simplify the computation, we use the following trick. 
We evaluate \eqref{eq:Veff-2-order-impl} in the IR regime, expanding for $k^2\ll M^2$, and retain terms up to $\mathcal O(k^2)$. Doing so, we are sure to keep terms at most linear in $R$. Starting at $\mathcal O(k^4)$ there are higher-curvature contributions too. Also, such an approximation``localizes" the form factor of $R\ph^2$ in the sense that only its constant ($k$-independent) part survives.

This is a simplification that significantly reduces the amount of calculation and, because of that, loses a great amount of information stored in \eqref{eq:Veff-2-order-impl}. However, it is enough for a cross-check between the $\xi$ sector of the effective action (that we derived in the section above) and the $\xi$ sector of the effective potential we are deriving now.

	After applying the IR limit, the methodology is the same as before. Details on the calculations are reported in Appendix \ref{App:eff-pot}. Here we report the final result,
	\begin{align}\label{2potential}
		V_{eff}^{(2)} &= \int d^4x  \left\{\left(- \frac{1}{4}h_{\mu\nu}^2 + \frac{1}{8}h^2 \right)\left[\frac{1}{2}m^2\ph^2+\frac{\lambda}{4!}\ph^4\right]+\frac{1}{2}h\left(\frac12\xi\ph^2R^{(1)}\right)+\left(\frac12\xi\ph^2R^{(2)}\right)\right\}\nn\\
		&+\frac12 \int d^4x \Bigg\{\left(- \frac{1}{4}h_{\mu\nu}^2 + \frac{1}{8}h^2 \right)\frac{M^4}{2(4\pi)^2} (\mu^2)^{\omega-2} \left[ -\frac{1}{2-\omega} - \frac{3}{2} + \ga+\ln\left(\frac{M^2}{4\pi\mu^2}\right) \right]\nn\\
		&+\frac{M^2}{(4\pi)^2} (\mu^2)^{\om-2}\left(\xi-\frac16\right)\left[ -\frac{1}{2-\omega} - 1 + \ga+\ln\left(\frac{M^2}{4\pi\mu^2}\right) \right]\left(\frac12hR^{(1)} + R^{(2)}\right)\Bigg\}\,.
	\end{align}

	An analysis of the previous orders reveals a consistent structural pattern dictated by general covariance. The terms proportional to $M^4$ and $M^2$ systematically reconstruct the covariant invariants $\sqrt{g}$ and $\sqrt{g}R$, respectively. Furthermore, these structures are consistent with the requirement that the divergences must respect conformal symmetry in the limit of a massless, conformally coupled scalar field ($m^2=0$ and $\xi=1/6$). Consequently, upon resumming all orders of the metric fluctuation $h_{\mu\nu}$, the complete unrenormalized effective potential is expected to be
	\begin{align}
		V_{eff} \, = \, &\int d^4x \sqrt{g}\left\{\frac{1}{2}m^2\phi^2+\frac{\la}{4!}\phi^4+\frac{1}{2}\xi R\phi^2\right\}\nn\\
		&+\frac12\int d^4x\sqrt{g}\bigg\{\frac{M^4}{2(4\pi)^2} (\mu^2)^{\omega-2} \left[ -\frac{1}{2-\omega} - \frac{3}{2} + \ga+\ln\left(\frac{M^2}{4\pi\mu^2}\right)\right]\nn\\
		&+\frac{M^2}{(4\pi)^2} (\mu^2)^{\om-2}\left(\xi-\frac16\right) \left[ -\frac{1}{2-\omega} - 1 + \ga+\ln\left(\frac{M^2}{4\pi\mu^2}\right) \right]R\bigg\}\,.
	\end{align}

	It is worth noting that a simpler approach for performing these
    calculations is to consider the trace mode of $h_{\mu\nu}$ only,
    while setting the transverse and traceless (TT) part to zero.
    The TT and trace parts capture the spin-2 and spin-0 components,
    respectively. Since our primary interest lies in the one-loop
    correction to the non-minimal coupling $\xi R\ph^2$, one could isolate the conformal mode from the
    beginning by substituting
    $h_{\mu\nu}\to\delta_{\mu\nu}\frac{h}{2\omega}$.
    This projection would drastically simplify the tensor calculations.
    However, by performing the calculation with the full metric
    perturbation $h_{\mu\nu}$, we demonstrate that the
    separate tensor contributions recombine into the $R\ph^2$ term,
    explicitly reconstructing the full covariant Ricci scalar without
    restricting the metric fluctuation to a purely conformal mode
    a priori.

    We continue by discussing the renormalization of the effective
potential. After introducing appropriate counterterms to absorb
the poles generated by dimensional regularization, the renormalized
effective potential in four dimensions can be parametrized, up to
$\ph$-independent terms, as
\begin{align} \label{eq:Veff-Ren}
    V_{eff}^R \, = \, \int d^4x \sqrt{g}\,\mathcal{V}_{\mathrm{eff}}^R
    \, = \, {}&
    \int d^4x \sqrt{g}
    \bigg\{
    \frac{1}{2}m_{\mathrm{eff}}^2(\ph)\ph^2
    +\frac{\la_{\mathrm{eff}}(\ph)}{4!}\ph^4
    +\frac{1}{2}\xi_{\mathrm{eff}}(\ph)R\ph^2
    \bigg\},
\end{align}
with
\begin{align}\label{RG}
    m_{\mathrm{eff}}^2(\ph)
    &=
    m^2+\frac{\la m^2}{2(4\pi)^2}
    \left[
    -\frac{3}{2}
    +\ln\left(
    \frac{m^2+\frac{\la}{2}\ph^2}{\bar\mu^2}
    \right)
    \right],
    \nn\\
    \la_{\mathrm{eff}}(\ph)
    &=
    \la+\frac{3\la^2}{2(4\pi)^2}
    \left[
    -\frac{3}{2}
    +\ln\left(
    \frac{m^2+\frac{\la}{2}\ph^2}{\bar\mu^2}
    \right)
    \right],
    \nn\\
    \xi_{\mathrm{eff}}(\ph)
    &=
    \xi+\frac{\la}{2(4\pi)^2}
    \left(\xi-\frac16\right)
    \left[
    -1
    +\ln\left(
    \frac{m^2+\frac{\la}{2}\ph^2}{\bar\mu^2}
    \right)
    \right].
\end{align}
These quantities are field-dependent effective parameters and should
not be confused with the field-independent renormalized couplings
defined at a fixed normalization point.
	In eq.~\eqref{RG}, we consider the $\overline{\mathrm{MS}}$ scheme, in
which the Euler--Mascheroni constant $\gamma$ and the factor
$\ln(4\pi)$ are absorbed into the definition of the renormalization
scale, and we have substituted the shifted mass $M^2$. The finite
constants $-3/2$ and $-1$ cannot be simultaneously
eliminated by a single rescaling of $\mu$. These constants multiply
local terms and depend on the choice of renormalization prescription.
Consequently, they may be shifted independently by adding appropriate
finite local counterterms.

Since these finite constants carry no explicit dependence on either
the renormalization scale or the subtraction momentum, they do not
contribute to the corresponding beta functions. 
Accordingly, our
result \eqref{eq:Veff-Ren} agrees with that of ref.~\cite{BuchbinderShapiro:2021} up to
finite local, scheme-dependent terms.

Let us conclude by commenting on the comparison between
the one-loop effective action and the effective potential. The two results must be equivalent in the following limits. The one-loop effective action, $\bar{\Ga}^{(1)}$, must be evaluated in the IR regime at $\cl{O}(k^2)$ precision, and for a constant $\ph$, 
\begin{align}\label{eq:effective-action-IR}
    \bar{\Ga}^{(1)}\Big|_{\text{IR}}
    =
    -\frac{1}{2(4\pi)^2}
    \int\dd[4]{x}\sqrt{g}\,
    \frac{\la}{2}\varphi^2
    \qty(\xi-\frac16)
    \left[
        \frac{1}{2-\om}
        -\ga
        -\ln\qty(\frac{m^2}{4\pi\mu^2})
    \right]R\,.
\end{align}
On the other hand, the effective potential must be expanded for $\frac{\la}{2}\ph^2\ll m^2$, which ensures the perturbative approach for small $\lambda$. This limit is not implemented when considering the shifted mass $M$. Let us write the $\xi$ contribution to the effective potential (taking $\om=2$ where possible),
\begin{align}
	V_{\mathrm{eff}}
    &=
    \frac{1}{2(4\pi)^2}
    \int d^4x \sqrt{g}
    \bigg\{
        \left(m^2+\frac{\la}{2}\varphi^2\right)
        \left(\xi-\frac16\right)
        \bigg[
            -\frac{1}{2-\omega}
            -1+\ga
            \nn\\
    &\qquad\qquad
            +\ln\left(\frac{m^2}{4\pi\mu^2}\right)
            +\ln\left(1+\frac{\la\varphi^2}{2m^2}\right)
        \bigg]R
    \bigg\}
    \nn\\
    &=
    \frac{1}{2(4\pi)^2}
    \int d^4x \sqrt{g}
    \bigg\{
        \frac{\la}{2}\varphi^2
        \left(\xi-\frac16\right)
        \left[
            -\frac{1}{2-\om}
            +\ga
            +\ln\left(\frac{m^2}{4\pi\mu^2}\right)
        \right]R
        \nn\\
    &\qquad\qquad
        +\text{terms proportional to }m^2R
        +\mathcal{O}(\varphi^4R)
    \bigg\}.
\end{align}
Here, we used the expansion of the logarithm around $\varphi=0$ equivalent to the small $\lambda$ expansion.
The two results are indeed equivalent.

\section{Conclusions}
\label{sec:5}
	The one-loop effective action is a fundamental quantity in quantum field theory, carrying the corrections to the dynamics of a classical background field. In flat space-time, such a quantity is commonly derived via Feynman diagrams, providing predictions with astonishing precision. On the other hand, in curved space-time, the need to preserve covariance has pushed the development of more elaborate methods, such as the heat-kernel and Riemann normal coordinates.
	
	In this work, we have recovered the intuition of Feynman diagrams by using the linearized metric approach in curved spacetime, where we treat gravity as an external field and only quantize the matter sector. 
	From the computational point of view, this choice is highly advantageous, since the expansion around a flat background enables one to use the Fourier transform to solve the loop integrals in momentum space.
	
	By using the linear parametrization of the metric and working order by order in $h_{\mu\nu}$, we break the manifest covariance of the theory. However, this is the core of this method, based on what we call the "principle of covariance": it is expected that the exact covariant structure can be inferred from the first orders of the expansion, since summing up infinitely many diagrams will restore full covariance. Therefore, by computing just the first orders, one can understand what the covariant structure of the quantum corrections is.
	
	We considered a real scalar field theory with quartic self-interaction and derived the one-loop quantum correction to the $\xi R\phi^2$-type term. We did so in linearized gravity, computing the diagram in fig.\ \ref{fig:bomb-diagram}. This is the only diagram with an $h_{\mu\nu}$ leg contributing to this type of term and inducing the running of the corresponding coupling $\xi$. From the loop result, its tensor structure allows one to reconstruct the scalar curvature at the first order in perturbation. Such a result perfectly illustrates the covariance principle: if we sum the contribution of this diagram with the other tadpoles (inserting more external metric legs), we would obtain the complete expansion of the scalar curvature.
	
	The loop calculation was done by using dimensional regularization, which enabled us to find the finite part of the loop and study the behavior of its form factor in the IR and UV regimes. From this result, using the momentum subtraction (MOM) scheme, we explored the running of the non-minimal coupling $\xi$ and its behavior in the IR and UV. Overall, the $\xi$ sector of the one-loop effective action induced by the loop in fig. \ref{fig:bomb-diagram} is in agreement with the effective action derived by the covariant method of heat kernel \cite{deBerredo-Peixoto:2003pkj}.
	
	Finally, to show the power and versatility of the method, we have computed the one-loop effective potential of this theory. From renormalizability, one should obtain counterterms proportional to $\ph^4$ and $\ph^2$. For the first case, expanding $\text{Tr} \log H$ at first order in $h_{\mu\nu}$ is sufficient to capture the coefficient in front of this counterterm. However, for the second term (the part proportional to the scalar curvature), its contribution can be understood only by calculating second-order diagrams. This occurs because, at the first order, the scalar curvature is a boundary term and its integration in coordinate space vanishes when we consider $\ph$ constant.
	 in the end, after the calculation of the corresponding diagrams \ref{fig:effective-potential-orders}, we verified that our result is in agreement with the effective potential derived from the effective action \cite{deBerredo-Peixoto:2003pkj}.
	We also commented on the agreement between our renormalized effective potential and the one derived directly with Riemann normal coordinates \cite{BuchbinderShapiro:2021}, up to renormalization scheme constants.
	 
	From the mathematical side, it seems possible to generalize the linearized gravity approach to theories with other fields and higher loops. For example, for additional real scalar fields or a complex scalar field, the procedure is straightforward. For extensions with potentially interesting applications, such as gauge theories and fermions, one faces a greater algebraic complexity. However, one can apply the Faddeev-Popov method and construct the ghost Hessian before expanding the metric. In this way, the main difficulty is restricted to the explicit expansion of covariant derivatives, tetrads, and spin connections. Although this requires additional computational effort, the procedure remains theoretically consistent.

\section*{Acknowledgments}
	The authors acknowledge Thomas M. Sangy for his involvement in the early stages of the project and are grateful to Ilya L. Shapiro and Gustavo P. de Brito for their support and insightful discussions. The authors acknowledge Alícia G. Borges for useful discussions.
	The authors are grateful to Coordenação de Aperfeiçoamento de Pessoal de Nível Superior (CAPES) for supporting their Ph.D.\ projects.

\newpage
\appendix

\section{Useful expansions}
\label{app:A1}
We list the expansion of useful quantities in orders of $h_{\mu\nu}$, for a metric parametrization $g_{\mu\nu}=\delta_{\mu\nu}+h_{\mu\nu}$.
We start from the inverse metric
\begin{align}
	g^{\mu\nu} = \delta^{\mu\nu} - h^{\mu\nu} + h^{\mu\tau}h_\tau{}^\nu +\mathcal{O}(h^3) 
\end{align}
used for expanding the box operator and the Ricci curvature,
\begin{align}
		&\square = \partial^2 - h^{\mu\nu}\partial_\mu\partial_\nu - (\partial_\mu h^{\mu\tau})\partial_\tau + \frac{1}{2}(\partial^\tau h)\partial_\tau +  h^{\mu\nu} h_\mu^{\,\,\alpha} \partial_\alpha \partial_\nu - \frac12h^{\mu\nu} (\partial_\la h_{\mu\nu})\partial^\la\nn\\
		&\quad-\, \frac12 h_{\mu\nu} (\partial^\nu h)\partial^\mu +h^{\nu\alpha} (\partial_\alpha h_{\mu\nu})\partial^\mu+  h_\mu^{\,\,\nu} (\partial_\alpha h_\nu^{\,\,\alpha})\partial^\mu+ \mathcal{O}(h^3) \nn \,,\\
		&R = \partial_\mu\partial_\nu h^{\mu\nu} - \partial^2 h + h^{\mu\nu} \partial_\mu \partial_\nu h - 2 h^{\mu\nu} \partial_\nu \partial_\alpha h_\mu^{\,\,\alpha}  +  h^{\mu\nu} \partial^2 h_{\mu\nu} - \frac{1}{4} (\partial^\nu h) (\partial_\nu h) \nonumber \\
		&-  (\partial_\mu h^{\mu\nu}) (\partial_\alpha h_{\,\,\nu}^\alpha)+   (\partial^\nu h) (\partial_\alpha h_{\,\,\nu}^\alpha) -\frac12 (\partial_\nu h_{\mu\alpha}) (\partial^\alpha h^{\mu\nu}) + \frac{3}{4} (\partial_\alpha h_{\mu\nu}) (\partial^\alpha h^{\mu\nu}) +\mathcal{O}(h^3)\,,  \label{eq:expansion-Ricci-scalar-flat-background}
\end{align}
where $\partial^{2} = \delta^{\mu\nu}\partial_{\mu}\partial_{\nu}$ and $h = \delta^{\mu\nu} h_{\mu\nu}$. 
Then, the expansion of the determinant of the metric is
\begin{align}
		\sqrt{g} = 1 + \frac{1}{2}h - \frac{1}{4}h_{\mu\nu}^2 + \frac{1}{8}h^2 +\mathcal{O}(h^3)\,. 
\end{align}

\section{Loops in dimensional regularization} \label{app:A2}
	The common procedure to compute loops such as \eqref{eq:loop-bomb-diagram-momentum-space} is to use the Feynman formula,
	\begin{align}
		\frac{1}{ab} \,=\, \int_0^1\frac{d\al}{\left[a \al+b(1-\al)\right]^2}, 
	\end{align}
	which can be easily demonstrated. This is a trick to rewrite the denominator as an even function of the momentum so that symmetry arguments can be used when performing the integral. 
	Choosing $a=(p+k)^2+m^2$ and $b=p^2+m^2$, the denominator of \eqref{eq:loop-bomb-diagram-momentum-space} reads
	\begin{align} \label{eq:Feynman-trick-denominator}
		& \frac{1}{\left(p^2+m^2\right)\left[(p+k)^2 + m^2\right]} = \int_0^1 \dd{\al} \frac{1}{[(p+\al k)^2+\Delta^2(k,\al)]^2} \nn\\
		& \text{with} \qquad \Delta^2=k^2\al(1-\al)+m^2 \,.
	\end{align}
	By applying a change of variable $p_\mu \to p_\mu -\al k_\mu$ we further simplify the denominator. After these steps, the loop \eqref{eq:loop-bomb-diagram-momentum-space} reads
	\begin{align} \label{eq:loop-integral-after-Feynman}
		I_{\mu\nu}(k) = I_{\mu\nu}^{(0)} + I_{\mu\nu}^{(1)} + I_{\mu\nu}^{(2)} \,, \qquad 
		I_{\mu\nu}^{(i)} = N(\la) \int_0^1 \dd{\al} \int\frac{\dd[4]{p}}{(2\pi)^{4}} \frac{f^{(i)}_{\mu\nu}(k,p^i)}{[p^2+\Delta^2]^2}
	\end{align}
	with
	\begin{align}
		& f^{(0)}_{\mu\nu}(k,\al) = \frac12 \delta_{\mu\nu} m^2 -\xi(k_\mu k_\nu -\delta_{\mu\nu} k^2) +\frac12 \delta_{\mu\nu}\al(\al-1)k^2 -\al(\al-1)k_\mu k_\nu \,, \label{eq:f0} \\
		& f^{(1)}_{\mu\nu}(k,p,\al) = \frac12 \delta_{\mu\nu} p.k -k_\mu p_\nu -\delta_{\mu\nu} \al k.p + \al( p_\mu k_\nu + p_\nu k_\mu) \,, \\
		& f^{(2)}_{\mu\nu}(k,p^2) = \frac12 \delta_{\mu\nu} p^2 -p_\mu p_\nu \,. \label{eq:f2}
	\end{align}
	The index $i$ denotes the power of the loop momentum in the numerator and functions $f_{\mu\nu}^{(i)}$ follow from the vertex structure $V_{\mu\nu}$ under the change of coordinates.
	
	We use dimensional regularization to deal with divergences, thus moving to dimension $2\om$ with $\om$ a complex number. Looking at \eqref{eq:loop-integral-after-Feynman} we identify three different integral structures,
	\begin{subequations} \label{eq:momentum-integral-basis}
		\begin{align}
			& J^{(0)}(k,\al) = \int\frac{\dd[2\om]{p}}{(2\pi)^{2\om}} \frac{1}{[p^2+\Delta^2]^2} \,, \\
			& J^{(1)}_\rho(k,\al) = \int\frac{\dd[2\om]{p}}{(2\pi)^{2\om}} \frac{p_\rho}{[p^2+\Delta^2]^2} \,, \\
			& J^{(2)}_{\rho\sigma}(k,\al) = \int\frac{\dd[2\om]{p}}{(2\pi)^{2\om}} \frac{p_\rho p_\sigma}{[p^2+\Delta^2]^2} \,.
		\end{align}
	\end{subequations}

	The integrals $J^{(i)}(k,\al)$ can be computed in spherical coordinates where, because of spherical symmetry, we can integrate the angular part obtaining 
	\begin{align}
		\dd[2\om]{p} = \frac{2 \pi^\omega}{\Ga(\om)} p^{2\om-1} \dd{p} \,.
	\end{align}
	Here $\Ga(z)$ is the Gamma-function. The integral is then performed by using the Beta-function
	\begin{align}
		B(x,y) = \frac{\Ga(x)\Ga(y)}{\Ga(x+y)} = \int_0^\infty \dd{t} t^{x-1}(1+t)^{-x-y}.
	\end{align}
	Quantities $J^{(i)}$ reduce to Beta-functions under the reparameterization $p^2=t\Delta^2$. Let us mention the Gamma function's property $\Ga(z+1)=z\Ga(z)$, which will be used in what follows.
	
	In more detail, the integral $J^{(0)}$ reduces to
	\begin{align} \label{eq:J0}
		J^{(0)} = \frac{\Ga(2-\om)}{(4\pi)^\om} \Delta^{2(\om-2)}
	\end{align}
	since $\Ga(2)=1$.
	
	Integral $J^{(1)}_\rho$ is zero for symmetry reasons. The integrand is an odd function integrated over a symmetric domain. 
	Integral $J^{(2)}_{\rho\si}$ reduces to
	\begin{align} \label{eq:J2}
		J^{(2)}_{\rho\sigma} = -\frac{\Ga(2-\om)}{2(4\pi)^\om (1-\om)} \Delta^{2(\om-1)} \delta_{\rho\si} =   -\frac{\Ga(1-\om)}{2(4\pi)^\om} \Delta^{2(\om-1)} \delta_{\rho\si}\,,
	\end{align}
	where we used the identity
	\begin{align}
		\int \dd[d]{p} \frac{p_\rho p_\si}{f(p^2)} = \frac{1}{d} \int \dd[d]{p} \frac{p^2\delta_{\rho\si}}{f(p^2)}
	\end{align}
	with $f(p^2)$ an even function. 
	
	The divergences of integrals \eqref{eq:momentum-integral-basis} are encoded into the Gamma-function $\Ga(2-\om)$. Indeed, $\Ga(z)$ has poles at $z=0$ and at negative integer arguments. Thus, in the limit $\om\to 2$ in which the four-dimensional spacetime is recovered, $\Ga(2-\om)$ diverges.
	The purpose of regularization methods is to parameterize divergences, thus it is reasonable to expand the integrals $I^{(i)}$ to $\cl{O}(2-\om)$ because these contributions vanish in the limit $\om\to 2$.
	
	In doing so we use the following definition of the Gamma-function,
	\begin{align} \label{eq:def-limit-Gamma-function}
		\Ga(z) = \lim_{n\to \infty} \frac{n! n^z}{z(z+1)\dots (z+n)} \,,
	\end{align}
	and expand it for $z\to 0$ up to $\cl{O}(z)$. First of all, it is easy to see from \eqref{eq:def-limit-Gamma-function} that $\Ga(z) \sim 1/z$ for $z\to 0$. Then the expansion of the numerator and of the terms $(z+k)^{-1}$ must go up to $\cl{O}(z^2)$. In the numerator
	\begin{align}
		n^z = e^{z\ln n} = 1+z\ln n +\cl{O}(z^2) \,.
	\end{align}
	Instead in the denominator
	\begin{align}
		\frac{1}{(z+1)\dots (z+n)} 
		&= (1-z+\cl{O}(z^2)) \dots \frac{1}{n}\qty(1-\frac{z}{n} +\cl{O}(Z^2)) \nn \\
		&= \frac{1}{n!}\qty[ 1-z\qty(1+\frac12+\dots\frac{1}{n})] +\cl{O}(z^2) \,.
	\end{align}
	Putting everything together, 
	\begin{align} \label{eq:Gamma-function-expansion-zero}
		\Ga(z) = \frac{1}{z} -\gamma +\cl{O}(z) \qquad \text{for} \qquad z\to 0
	\end{align}
	where we introduced the Euler-Mascheroni constant,
	\begin{align}
		\lim_{n\to\infty} \qty[ 1+\frac12+\dots +\frac{1}{n} -\ln n ] = \gamma \,.
	\end{align}

	In order to properly expand the integrals \eqref{eq:J0} and \eqref{eq:J2} at $\cl{O}(2-\om)$ we have to expand the power of $\Delta$. Let us start from \eqref{eq:J0} where there is
	\begin{align}
		\frac{\Delta^{2(\om-2)}}{(4\pi)^{\om}} = \frac{(\mu^2)^{\om-2}}{(4\pi)^2} \left(\frac{\Delta^2}{4\pi\mu^2}\right)^{\om-2} =\frac{(\mu^2)^{\om-2}}{(4\pi)^2}\left[1+(\om-2)\ln\left(\frac{\Delta^2}{4\pi\mu^2}\right)+\mathcal{O}(\om-2)^2\right] \,.
	\end{align}
	We introduced a mass scale $\mu$ to make the argument of the logarithm dimensionless.
	Now we can put together all the terms entering the integral \eqref{eq:J0} to find
	\begin{align} \label{eq:J0-final}
		J^{(0)}(k,\al) = \frac{\mu^{2(\om-2)}}{(4\pi)^2} \qty[ \frac{1}{(2-\om)} -\ga -\ln\qty(\frac{\Delta^2(k,\al)}{4\pi\mu^2}) ] + \cl{O}(\om-2) \,.
	\end{align}
	The dimensionally regularized contribution from the $J^{(0)}$ to the loop integral \eqref{eq:loop-integral-after-Feynman} is therefore
	\begin{align} \label{eq:loop-integral-contibution-0}
		I^{(0)}_{\mu\nu}(k) &= N(\la) \int_0^1 \dd{\al} J^{(0)}(k,\al) f^{(0)}_{\mu\nu}(k,\al)\nn\\
        & = \frac{N(\la)\mu^{2(\om-2)}}{(4\pi)^2} \int_0^1\dd{\al}\qty[ \frac{1}{(2-\om)} -\ga -\ln\qty(\frac{\Delta^2(k,\al)}{4\pi\mu^2}) ]\times \nn\\
        &\qquad \times\left[\frac12 \delta_{\mu\nu} m^2 -\xi(k_\mu k_\nu -\delta_{\mu\nu} k^2) +\frac12 \delta_{\mu\nu}\al(\al-1)k^2 -\al(\al-1)k_\mu k_\nu\right] \,,
	\end{align}
	with $f^{(0)}$ defined in \eqref{eq:f0}.
	Thanks to the expansion in \eqref{eq:J0-final} it is easy to separate the divergent part from the finite one. 
	
	Before commenting on $I^{(0)}_{\mu\nu}(k)$ let us expand \eqref{eq:J2}, as done with \eqref{eq:J0}. The power of $\Delta^2$ can be written as
	\begin{align}
		\Delta^2 \Delta^{2(\om-2)} = \Delta^2 \mu^{2(\om-2)} \qty[ 1+(\om-2) \ln\qty(\frac{\Delta^2}{\mu^2})] +\cl{O}(\om-2) \,.
	\end{align}
	Then, using \eqref{eq:Gamma-function-expansion-zero}, the integral \eqref{eq:J2} reads
	\begin{align} \label{eq:J2-final}
		J^{(2)}_{\rho\si}(k,\al) = \frac{\Delta^2 \mu^{2(\om-2)}}{2(4\pi)^2} \delta_{\rho\si} \qty[ -\frac{1}{(2-\om)} -1+\ga +\ln\qty(\frac{\Delta^2(k,\al)}{4\pi\mu^2}) ] + \cl{O}(\om-2) \,.
	\end{align}
	And the corresponding contribution to the loop integral is
	\begin{align} \label{eq:loop-integral-contibution-2}
		I_{\mu\nu}^{(2)}(k) &= N(\la) \int_0^1\dd{\al} \qty[\frac12 \delta_{\mu\nu} \delta^{\rho\si} J_{\rho\si}^{(2)}(k,\al) - J_{\mu\nu}^{(2)}(k,\al)]\nn\\
        &=-\frac{N(\la) \mu^{2(\om-2)}}{2(4\pi)^2} \delta_{\mu\nu}\int_0^1\dd{\al} \Delta^2\left[\frac{1}{2-\om} - \ga -\ln\left(\frac{\Delta^2(k,\al)}{4\pi\mu^2}\right)+\mathcal{O}(2-\om)\right]
	\end{align}
	where we substituted $f^{(2)}_{\mu\nu}$ with its explicit expression \eqref{eq:f2}.

	\section{Divergent part}
	Eqs.\ \eqref{eq:loop-integral-contibution-0}, \eqref{eq:loop-integral-contibution-2} are the loop integrals in dimensional regularization. Let us separate the divergent part, proportional to the pole $1/(2-\om)$ and perform the $\al$-integration that, in the divergent sector, is straightforward.
	It reduces to integrating $f^{(0)}_{\mu\nu}$ defined in \eqref{eq:f0} or $\Delta^2$ defined in \eqref{eq:Feynman-trick-denominator}.
	Then
	\begin{align}
		I_{\mu\nu}^{\text{div}}(k) = I_{\mu\nu}^{(0)}{}^{\text{div}} + I_{\mu\nu}^{(2)}{}^{\text{div}}
	\end{align}
	with
	\begin{align}
		& I_{\mu\nu}^{(0)}{}^{\text{div}}(k) = \frac{N(\la)}{\varepsilon} \qty[\frac12 \delta_{\mu\nu} m^2 -\xi(k_\mu k_\nu -\delta_{\mu\nu} k^2) -\frac16\qty( \frac12 \delta_{\mu\nu} k^2 -k_\mu k_\nu)] \,, \\
		& I_{\mu\nu}^{(2)}{}^{\text{div}}(k) = -\frac{N(\la)}{2\varepsilon} \qty[\frac{k^2}{6} + m^2] \delta_{\mu\nu} \,,
	\end{align}
	where
	\begin{align} \label{eq:pole-definition-varepsilon}
		\frac{1}{\varepsilon} \equiv \frac{\mu^{2(\om-2)}}{(4\pi)^2 (2-\om)} \,.
	\end{align}
	Summing together, the divergent part of the diagram \ref{fig:bomb-diagram} is
	\begin{align} \label{eq:divergent-part-diagram}
		I_{\mu\nu}^{\text{div}}(k) = \frac{N(\la)}{\varepsilon} \qty(\xi-\frac16) (\delta_{\mu\nu} k^2-k_\mu k_\nu)\,.
	\end{align}

	Let us conclude by mentioning that integrals $J^{(i)}$ with $i>2$ can be found in \cite{Gorbar:2003I}, where one-loop scalar-field corrections to the graviton propagator were computed. Because General Relativity is a second-order-in-derivatives theory, the vertices considered in \cite{Gorbar:2003I} contain integrals $I^{(i)}$ with $i>2$.

	\section{Finite part} \label{App:finite-part}
	Let us carry out the computation of the finite part of the diagram \ref{fig:bomb-diagram} starting from the regularized expressions \eqref{eq:loop-integral-contibution-0} and \eqref{eq:loop-integral-contibution-2}. The finite part will be given by
	\begin{align}
		I_{\mu\nu}^{\text{fin}}(k) = I_{\mu\nu}^{(0)}{}^{\text{fin}} + I_{\mu\nu}^{(2)}{}^{\text{fin}}
	\end{align}
	where
	\begin{align}
		& I_{\mu\nu}^{(0)}{}^{\text{fin}}(k) = -N(\la)\frac{\mu^{2(\om-2)}}{(4\pi)^2} \int_0^1\dd{\al} \qty[\ga +\ln\qty(\frac{\Delta^2(k,\al)}{4\pi\mu^2})]f^{(0)}_{\mu\nu}(k,\al) \,, \label{eq:finite-part-0}\\
		& I_{\mu\nu}^{(2)}{}^{\text{fin}}(k) = \frac{N(\la)}{2} \frac{\mu^{2(\om-2)}}{(4\pi)^2} \int_0^1\dd{\al}  \Delta^2(k,\al)\qty[ \ga +\ln\qty(\frac{\Delta^2(k,\al)}{4\pi\mu^2}) ] \delta_{\mu\nu} \label{eq:finite-part-2}
	\end{align}
	The above expressions are obtained by isolating the finite part of \eqref{eq:J0-final} and \eqref{eq:J2-final}. Let us mention that in the second line we wrote the trace $\delta^\rho{}_\rho=2\om$.
	We also wrote explicitly the dependence on $\al$ to stress which integrals need to be computed. Indeed, in the finite sector, the $\al$-integrals are more involved than the ones in the divergent sector. Finally, we recall that $f^{(0)}_{\mu\nu}$ was defined in \eqref{eq:f0}, while $\Delta^2$ in \eqref{eq:Feynman-trick-denominator}.
	
	Before proceeding with the integration let us rewrite the logarithm making the $\al$-dependence explicit,
	\begin{align}
		\log\qty(\frac{\Delta^2}{4\pi\mu^2}) = \log\qty(\frac{m^2}{4\pi\mu^2}) + \log\qty[ \al(1-\al)\tau + 1] 
	\end{align}
	where we introduced the dimensionless variable
	\begin{align}
		\tau \equiv \frac{k^2}{m^2}
	\end{align}
	 Note that in Euclidean signature it is a positive quantity. 
	
	Altogether, the $\al$-integrals that appear in the finite part of the loop consist of polynomials, coming from the $\al(1-\al)$ term, and two non-trivial integrals
	\begin{align} \label{eq:alpha-integrals}
		& \int_0^1\dd{\al} \log\qty[ \al(1-\al)\tau + 1] \equiv F_1(\tau) \,, \\
		& \int_0^1\dd{\al} \al(\al-1)\log\qty[ \al(1-\al)\tau + 1] \equiv F_2(\tau) \,.
	\end{align}
	We postpone the task of explicitly solving the integrals; for now we will use the compact notations $F_1$ and $F_2$ and keep manipulating $I_{\mu\nu}^{\text{fin}}(k)$.
	
	Let us start with \eqref{eq:finite-part-0}. In terms of $F_1,F_2$ it is easy to obtain
	\begin{align}
		&I_{\mu\nu}^{(0)}{}^{\text{fin}}(k) = -N(\la)\frac{\mu^{2(\om-2)}}{(4\pi)^2} \Big\{\qty[\qty(\ga +\ln\qty(\frac{m^2}{4\pi\mu^2})) \qty(\frac{m^2}{2} -\frac{k^2}{12}) +\frac{m^2}{2}F_1 +\frac{k^2}{2}F_2 ]\delta_{\mu\nu} \nn\\
		&\hspace{4cm} +\qty[\frac16 \qty(\ga +\ln\qty(\frac{m^2}{4\pi\mu^2})) -F_2]k_\mu k_\nu \nn\\
		&\hspace{4cm} -\qty[\qty(\ga +\ln\qty(\frac{m^2}{4\pi\mu^2})) +F_1]\xi(k_\mu k_\nu -\delta_{\mu\nu}k^2 ) \Big\} \,.
	\end{align}
	We now consider \eqref{eq:finite-part-2}, which reduces to
	\begin{align}
		&I_{\mu\nu}^{(2)}{}^{\text{fin}}(k) = N(\la)\frac{\mu^{2(\om-2)}}{(4\pi)^2} \left\{ \qty(\ga+\ln\qty(\frac{m^2}{4\pi\mu^2}))\qty[\frac{k^2}{12}+\frac{m^2}{2}]+\frac{m^2}{2}F_1 -\frac{k^2}{2}F_2  \right\} \delta_{\mu\nu} \,.
	\end{align}
	We can now sum the two contributions to the finite part of the diagram. After some easy algebra one finds
	\begin{align} \label{eq:finite-part-loop}
		I_{\mu\nu}^{\text{fin}}(k) =  \frac{N(\la)}{(4\pi)^2} \left\{ \qty[\qty(\xi-\frac16)\qty(\ga+\ln\qty(\frac{m^2}{4\pi\mu^2})) +\xi F_1(\tau) + F_2(\tau)] \qty[k_\mu k_\nu -\delta_{\mu\nu} k^2]\right\}.
	\end{align}
	In the expression above, the limit $\om\to2$ has been safely
    taken after separating the divergent part. An explicit expression for the quantity $\xi F_1+F_2$ is obtained once we solve the integrals \eqref{eq:alpha-integrals}, which is the next task.
	
	For both integrals \eqref{eq:alpha-integrals}, the procedure is to apply a change of variable $z=1/2-\al$ and then, after some manipulations, one finds\footnote{Here we are using the notation of \cite{BuchbinderShapiro:2021}.}
	\begin{align}
		F_1(a) = -2 Y \,, \qquad F_2= -\frac{1}{18} +\qty(\frac12 -\frac{2}{3a^2})Y \,,
	\end{align}
	where
	\begin{align} \label{eq:Y-a-definition}
		Y \equiv 1-\frac{1}{a}\ln\qty(\frac{2+a}{2-a}) \,, \quad a^2\equiv \frac{4\tau}{4+\tau} \,.
	\end{align}

	It follows that the quantity $\xi F_1+F_2$ explicitly reads
	\begin{align} \label{eq:F1+F2-form-factors}
		\xi F_1(a)+F_2(a) = -\frac{1}{18} -\qty(2\xi -\frac12 +\frac{2}{3a^2}) Y(a) \,.
	\end{align}

	\section{Integrals for the Effective Potential} \label{App:eff-pot} 
	
	In this appendix, we list the fundamental dimensional regularization integrals in Euclidean space required for the evaluation of the effective potential. We work in $2\omega$ dimensions and introduce the arbitrary mass scale $\mu$ to keep the coupling constants dimensionless.
	
	For the zeroth-order contribution to the effective potential, we encounter the momentum integral of the logarithm. Its evaluation can be obtained by noting that
	\begin{equation}
		\frac{\pa}{\pa M^2}\int\frac{d^{2\om}p}{(2\pi)^{2\om}}\log\left(p^2+M^2\right) = \int\frac{d^{2\om}p}{(2\pi)^{2\om}} \frac{1}{p^2+M^2}\,.
	\end{equation}
	The latter integral is well known,
	\begin{align} \label{eq:app_scalar}
		\int \frac{d^{2\om}p}{(2\pi)^{2\om}}\frac{1}{p^2+M^2} &= \frac{M^2}{(4\pi)^2} (\mu^2)^{\om-2}\left[ -\frac{1}{2-\omega} - 1 + \gamma + \ln\left( \frac{M^2}{4\pi\mu^2} \right) \right]\,.
		\end{align}
		Therefore, integrating with respect to $M^2$ we obtain the required logarithmic integral
	\begin{equation}
		\int\frac{d^{2\om}p}{(2\pi)^{2\om}}\log\left(p^2+M^2\right) = \frac{M^4}{2(4\pi)^2}(\mu^2)^{\om-2} \left[ -\frac{1}{2-\omega} - \frac{3}{2} + \gamma + \ln\left( \frac{M^2}{4\pi\mu^2} \right) \right]\,.
	\end{equation}
	For the first-order contribution we start by using the cyclicity of the trace,
\begin{align}
    \frac12\Tr G^{(0)}H^{(1)} = \frac12 \int d^4x \lim_{y\to x}\left[H^{(1)}(y)G^{(0)}(x,y)\right] \,.
\end{align}
Then we write $H_1$ and $G_0$ in momentum space, 
\begin{align}
    \frac12\Tr& G^{(0)}H^{(1)} = \frac12 \int d^4x \Bigg[\frac12 h(x)\delta^{4}(0)+\nn\\&+\int \frac{d^4k}{(2\pi)^4} \tilde{h}^{\mu\nu}(k) e^{-ikx} \int \frac{d^4p}{(2\pi)^4} \frac{-p_\mu p_\nu -k_\mu p_\nu + \frac12 \delta_{\mu\nu}p.k + \xi(k^2\delta_{\mu\nu}-k_\mu k_\nu)}{p^2+M^2}\Bigg]\,.
\end{align}
The integrals are evaluated in dimensional regularization, where it turns out that $\delta^{(2\om)}(0)= 0$, thus we can neglect the term proportional to $\delta^{(4)}(0)$.
In addition to this simplification, due to the parity of the integral, the terms with a numerator linear in $p_\mu$ vanish. It remains for us to evaluate the following integrals,
\begin{align}
    \int \frac{d^{2\om}p}{(2\pi)^{2\om}}\frac{1}{p^2+M^2} \qquad\text{ and }\qquad \int \frac{d^{2\om}p}{(2\pi)^{2\om}}\frac{p_\mu p_\nu}{p^2+M^2}\,.
\end{align}
The first integral is \eqref{eq:app_scalar}. The second integral follows easily
\begin{align}
		\int \frac{d^{2\om}p}{(2\pi)^{2\om}}\frac{p_\mu p_\nu}{p^2+M^2} &= -\frac{M^4\delta_{\mu\nu}}{4(4\pi)^2} (\mu^2)^{\omega-2} \left[ -\frac{1}{2-\omega} - \frac{3}{2} + \gamma + \ln\left( \frac{M^2}{4\pi\mu^2} \right) \right]\,.
\end{align}
Summing everything together one finds \eqref{1-correction}.

The second-order contribution to the effective potential \eqref{eq:Veff-2-order-impl} is, in coordinate space, 
\begin{align}
    &- \frac{1}{4}\Tr G^{(0)}H^{(1)}G^{(0)}H^{(1)}+\frac12\Tr G^{(0)}H^{(2)} = \nn\\
    &\quad=\frac12\int d^4x \left(-\frac{1}{4}h_{\mu\nu}^2(x)\delta^{(4)}(0)\right) +\frac12\int d^4x\Bigg[\left(-\Box^{(2)}+ \xi R^{(2)}\right)_x G^{(0)}(x,x)\Bigg]\nn\\
    &\quad-\frac14\int d^4x\int d^4z\left[\left(-\Box^{(1)}+ \xi R^{(1)}\right)_zG^{(0)}(x,z)\left(-\Box^{(1)}+ \xi R^{(1)}\right)_xG^{(0)}(z,x)\right]
\end{align}
Of these three integrals, we only need to calculate the last two, since the first vanishes in dimensional regularization. Thus, we must rewrite the last two integrals in the momentum representation. Let us start from the last one, 
\begin{align}
    I_1 = \frac14\int d^4x\int d^4z&\left[\left(-\Box^{(1)}+ \xi R^{(1)}\right)_zG^{(0)}(x,z)\left(-\Box^{(1)}+ \xi R^{(1)}\right)_xG^{(0)}(z,x)\right] = \nn\\
    &\hspace{-1cm}=\frac14\int d^4x\int_{r,k} \tilde h^{\mu\nu}(r) \tilde h^{\al\be}(k)e^{-i(k+r)x}\int_p \frac{V_{\mu\nu}[p,k]V_{\al\be}[p-k,r]}{(p^2+M^2)((p-k)^2+M^2)}\,, \nn\\
    &=\frac14\int_{k} \tilde h^{\mu\nu}(-k) \tilde h^{\al\be}(k)\int_p \frac{V_{\mu\nu}[p,k]V_{\al\be}[p-k,-k]}{(p^2+M^2)((p-k)^2+M^2)}\,,
\end{align}
with
\begin{align}
V_{\mu\nu}[p,k] =  -p_\mu p_\nu +\frac12(k_\mu p_\nu+k_\nu p_\mu) - \frac12 \delta_{\mu\nu}p.k + \xi(k^2\delta_{\mu\nu}-k_\mu k_\nu)
\end{align}
and we use the compact notation for the integral
\begin{align}
\int_{p} =\int \frac{d^4p}{(2\pi)^4}\,.
\end{align}
Furthermore, the other integral is
\begin{align}
I_2 = \frac12\int d^4x\Bigg[\left(-\Box^{(2)}+ \xi R^{(2)}\right)_x G^{(0)}(x,x)\Bigg] =\frac12\int_{k} \tilde h^{\mu\nu}(-k) \tilde h^{\al\be}(k)\int_p \frac{\mathcal{N}_{\mu\nu\al\be}[p,k]}{p^2+M^2}\,,
\end{align}
where
\begin{align}
\mathcal{N}_{\mu\nu\alpha\beta} &= \frac{1}{4} \Big( g^{\mu\alpha}p^\nu p^\beta + g^{\nu\alpha}p^\mu p^\beta + g^{\mu\beta}p^\nu p^\alpha + g^{\nu\beta}p^\mu p^\alpha \Big) \nonumber \\
&+\frac{\xi}{8} \Big[g^{\mu\alpha}k^\nu k^\beta + g^{\nu\alpha}k^\mu k^\beta + g^{\mu\beta}k^\nu k^\alpha + g^{\nu\beta}k^\mu k^\alpha-k^2 \Big( g^{\mu\alpha} g^{\nu\beta} + g^{\mu\beta} g^{\nu\alpha} + 2 g^{\mu\nu} g^{\alpha\beta} \Big)\Big]\nn\\
&+\frac{1}{8} \Big[ g^{\mu\nu} (p^\alpha k^\beta + p^\beta k^\alpha) - g^{\alpha\beta} (p^\mu k^\nu + p^\nu k^\mu) \Big]\,.
\end{align}

We expand the expressions for $I_1$ and $I_2$ in the low-momentum regime $k^2\ll M^2$,
to precision $\mathcal O(k^2)$. This is an approximation needed to isolate local terms linear in $\sqrt{g}R$, as explained in the main text.
The momentum integrals are then performed using dimensional regularization.  Definitions and results for these type of integrals are already stated in the appendices. We obtain
\begin{align}
    \left(I_2-I_1\right)_{\mathrm{div}} &= \frac{1}{4(4\pi)^2(2-\om)}\int_k \bigg\{ \frac{M^4}{8}\left[2\tilde h_{\mu\nu}(-k)\tilde h^{\mu\nu}(k)-\tilde h(-k)\tilde h(k)\right] \nn\\
    &\quad+\frac{M^2}{2}\left(\xi-\frac16\right)\Big[ 2 k^\mu k^\nu \tilde{h}(-k)\tilde{h}_{\mu\nu}(k) - 2 k^\mu k^\nu \tilde{h}_{\nu}^{\,\,\,\alpha}(-k) \tilde{h}_{\mu\alpha}(k) \nn\\
    &\quad\quad + k^2 \tilde{h}_{\mu\nu}(-k)\tilde{h}^{\mu\nu}(k) - k^2 \tilde{h}(-k)\tilde{h}(k) \Big]\bigg\}\nn\\
    &\hspace{-1cm}=\frac{1}{4(4\pi)^2(2-\om)}\int d^4x\bigg\{\left(\frac14 h_{\mu\nu}^2-\frac18h^2\right)M^4-2M^2\left(\xi-\frac16\right)\left(\frac12hR^{(1)} + R^{(2)}\right)\bigg\}.
\end{align}
The finite terms are obtained analogously in the same low-momentum expansion. Combining the divergent and finite contributions, we recover eq.~\eqref{2potential}.

\newpage

\printbibliography

@article{Coleman:1973jx,
    author = "Coleman, Sidney R. and Weinberg, Erick J.",
    title = "{Radiative Corrections as the Origin of Spontaneous Symmetry Breaking}",
    doi = "10.1103/PhysRevD.7.1888",
    journal = "Phys. Rev. D",
    volume = "7",
    pages = "1888--1910",
    year = "1973"
}

@article{deBerredo-Peixoto:2003pkj,
  author        = {de Berredo-Peixoto, Guilherme and Gorbar, Eduard V. and Shapiro, Ilya L.},
  title         = {On the renormalization group for the interacting massive scalar field theory in curved space},
  journal       = {Classical and Quantum Gravity},
  volume        = {21},
  number        = {9},
  pages         = {2281--2290},
  year          = {2004},
  doi           = {10.1088/0264-9381/21/9/005},
  eprint        = {hep-th/0311229},
  archiveprefix = {arXiv}
}

@article{Appelquist:1974tg,
  author        = {Appelquist, Thomas and Carazzone, J.},
  title         = {{Infrared Singularities and Massive Fields}},
  reportNumber  = {Print-74-1486 (HARVARD)},
  doi           = {10.1103/PhysRevD.11.2856},
  journal       = {Phys. Rev. D},
  volume        = {11},
  pages         = {2856},
  year          = {1975}
}

@book{BirrellDavies,
  author        = {Birrell, N. D. and Davies, P. C. W.},
  title         = {Quantum Fields in Curved Space},
  publisher     = {Cambridge University Press},
  year          = {1982}
}

@book{BuchbinderShapiro:2021,
  author        = {Buchbinder, Ilya L. and Shapiro, Ilya L.},
  title         = {Introduction to Quantum Field Theory with Applications to Quantum Gravity},
  publisher     = {Oxford University Press},
  year          = {2021},
  isbn          = {978-0198864899}
}

@article{Gorbar:2003I,
  author        = {Gorbar, Eduard V. and Shapiro, Ilya L.},
  title         = {Renormalization Group and Decoupling in Curved Space},
  journal       = {Journal of High Energy Physics},
  year          = {2003},
  volume        = {02},
  pages         = {021},
  doi           = {10.1088/1126-6708/2003/02/021},
  eprint        = {hep-ph/0210388},
  archiveprefix = {arXiv},
  primaryclass  = {hep-ph}
}

@article{Gorbar:2003II,
  author        = {Gorbar, Eduard V. and Shapiro, Ilya L.},
  title         = {{Renormalization group and decoupling in curved space. 2. The Standard model and beyond}},
  eprint        = {hep-ph/0303124},
  archivePrefix = {arXiv},
  reportNumber  = {DF-UFJF-03-01},
  doi           = {10.1088/1126-6708/2003/06/004},
  journal       = {JHEP},
  volume        = {06},
  pages         = {004},
  year          = {2003}
}

@article{Borges:2025gfg,
    author = "Borges, Al{\'\i}cia G. and Shapiro, Ilya L.",
    title = "{Derivative expansion in a two-scalar field theory}",
    eprint = "2503.23694",
    archivePrefix = "arXiv",
    primaryClass = "hep-th",
    doi = "10.1103/rynr-2znv",
    journal = "Phys. Rev. D",
    volume = "111",
    number = "12",
    pages = "125019",
    year = "2025"
}

@article{Franchino-Vinas:2018gzr,
  author        = {Franchino-Vi{\~n}as, Sebasti{\'a}n A. and de Paula Netto, Tib{\'e}rio and Shapiro, Ilya L. and Zanusso, Omar},
  title         = {{Form factors and decoupling of matter fields in four-dimensional gravity}},
  eprint        = {1812.00460},
  archivePrefix = {arXiv},
  primaryClass  = {hep-th},
  doi           = {10.1016/j.physletb.2019.01.021},
  journal       = {Phys. Lett. B},
  volume        = {790},
  pages         = {229--236},
  year          = {2019}
}

@article{Ribeiro:2018pyo,
  author        = {Ribeiro, Tiago G. and Shapiro, Ilya L. and Zanusso, Omar},
  title         = {{Gravitational form factors and decoupling in 2D}},
  eprint        = {1803.06948},
  archivePrefix = {arXiv},
  primaryClass  = {hep-th},
  doi           = {10.1016/j.physletb.2018.05.049},
  journal       = {Phys. Lett. B},
  volume        = {782},
  pages         = {324--331},
  year          = {2018}
}

@article{Buchbinder:1985ew,
  author        = {Buchbinder, I. L.},
  title         = {{Quantum Field Theory Renormalization in Curved Space-time and Renormalization Group Equations}},
  reportNumber  = {ITF-86-4P},
  doi           = {10.1002/prop.19860340902},
  journal       = {Fortsch. Phys.},
  volume        = {34},
  pages         = {605--628},
  year          = {1986}
}

@book{Peskin1995,
  author        = {Michael E. Peskin and Daniel V. Schroeder},
  title         = {An Introduction to Quantum Field Theory},
  publisher     = {Addison-Wesley},
  address       = {Reading, Massachusetts},
  year          = {1995},
  isbn          = {9780201503975}
}

@book{DeWitt:2003pm,
  author    = {DeWitt, Bryce S.},
  title     = {{The global approach to quantum field theory. Vol. 1, 2}},
  publisher = {Oxford University Press},
  volume    = {114},
  year      = {2003}
}

@article{Barvinsky:1985an,
  author        = {Barvinsky, A. O. and Vilkovisky, G. A.},
  title         = {{The Generalized Schwinger-Dewitt Technique in Gauge Theories and Quantum Gravity}},
  doi           = {10.1016/0370-1573(85)90148-6},
  journal       = {Phys. Rept.},
  volume        = {119},
  pages         = {1--74},
  year          = {1985}
}

@mastersthesis{Avramidi:1986mj,
  author        = {Avramidi, Ivan Grigorevich},
  title         = {{Covariant methods for the calculation of the effective action in quantum field theory and investigation of higher derivative quantum gravity}},
  school        = {Moscow State University},
  eprint        = {hep-th/9510140},
  archivePrefix = {arXiv},
  reportNumber  = {RX-1539},
  type          = {Other thesis},
  year          = {1986}
}

@article{Barvinsky:1990up,
    author = "Barvinsky, A. O. and Vilkovisky, G. A.",
    title = "{Covariant perturbation theory. 2: Second order in the curvature. General algorithms}",
    doi = "10.1016/0550-3213(90)90047-H",
    journal = "Nucl. Phys. B",
    volume = "333",
    pages = "471--511",
    year = "1990"
}

@misc{Barvinsky:2009CPT,
  author        = {Barvinsky, A. O. and Gusev, Yu. V. and
                   Zhytnikov, V. V. and Vilkovisky, G. A.},
  title         = {{Covariant Perturbation Theory (IV).
                    Third Order in the Curvature}},
  year          = {2009},
  eprint        = {0911.1168},
  archivePrefix = {arXiv},
  primaryClass  = {hep-th},
  doi           = {10.48550/arXiv.0911.1168},
  url           = {https://arxiv.org/abs/0911.1168},
  note          = {Originally circulated as report
                   PRINT-93-0274 (MANITOBA), 1993}
}

@misc{Mottola:2026vacuum,
  author        = {Mottola, Emil},
  title         = {{Gravitational Vacuum Polarization:
                    Decoupling and the Conformal Anomaly}},
  year          = {2026},
  eprint        = {2607.18180},
  archivePrefix = {arXiv},
  primaryClass  = {hep-th},
  url           = {https://arxiv.org/abs/2607.18180}
}
	
\end{document}